\documentclass[twocolumn, tighten]{aastex631} 
\usepackage{enumitem} 
\setlist[enumerate]{noitemsep,topsep=1ex}
\usepackage{graphicx}
\usepackage{amsmath}
\usepackage{amssymb}
\usepackage{textcomp}
\usepackage[caption=false]{subfig}
\usepackage{color}     
\usepackage{gensymb}   
\usepackage{hyperref}  
\usepackage{xspace}    
\usepackage{afterpage} 
\newcommand{\msun}{$\text{M}_{\odot}$\xspace}
\newcommand{\mstar}{$\text{M}_{\star}$\xspace}
\newcommand{\precover}{$P_\text{recover}$\xspace}
\shorttitle{}
\shortauthors{Garling et al.}
\begin{document}

\title{Measuring Resolved Star Formation Histories from High-Precision Color-Magnitude Diagrams with StarFormationHistories.jl}
\correspondingauthor{Christopher T. Garling}
\email{txa5ge@virginia.edu}

\author[0000-0001-9061-1697]{Christopher T. Garling}
\affiliation{Department of Astronomy, University of Virginia, 530 McCormick Road, Charlottesville, VA 22904, USA}

\author[0000-0002-3204-1742]{Nitya Kallivayalil}
\affiliation{Department of Astronomy, University of Virginia, 530 McCormick Road, Charlottesville, VA 22904, USA}

\author[0000-0001-5538-2614]{Kristen B. W. McQuinn}
\affiliation{Space Telescope Science Institute, 3700 San Martin Drive, Baltimore, MD 21218, USA}
\affiliation{Department of Physics and Astronomy, Rutgers, the State University of New Jersey,  136 Frelinghuysen Road, Piscataway, NJ 08854, USA}

\author[0000-0003-1634-4644]{Jack T. Warfield}
\affiliation{Department of Astronomy, University of Virginia, 530 McCormick Road, Charlottesville, VA 22904, USA}

\author[0000-0002-5581-2896]{Mario Gennaro}
\affiliation{Space Telescope Science Institute, 3700 San Martin Drive, Baltimore, MD 21218, USA}
\affiliation{The William H. Miller III Department of Physics \& Astronomy, Bloomberg Center for Physics and Astronomy, Johns Hopkins
University, 3400 N. Charles Street, Baltimore, MD 21218, USA}

\author[0000-0002-2970-7435]{Roger E. Cohen}
\affiliation{Department of Physics and Astronomy, Rutgers, the State University of New Jersey,  136 Frelinghuysen Road, Piscataway, NJ 08854, USA}



\begin{abstract}
Understanding how and when galaxies formed stars over the history of the Universe is fundamental to the study of galaxy evolution. The star formation histories (SFHs) of galaxies in the local Universe can be measured with high precision using deep imaging with space telescopes. Such \emph{resolved} SFHs are based on modelling the observed color-magnitude diagram (CMD) with stellar evolution models and rely on age-sensitive features like the main sequence turn-off to measure a galaxy's star formation rate as a function of time. There are many other population-level parameters that factor into these measurements, such as the stellar initial mass function (IMF), binary fraction, and metallicity, to name a few. We present and release \textbf{StarFormationHistories.jl}, a modular, open-source Julia package for measuring resolved SFHs with a focus on model flexibility for these types of population parameters. The code can model unresolved photometric binaries and supports arbitrary IMFs. Random uncertainties in the SFH measurements can be quantified with Monte Carlo posterior sampling methods. We illustrate the performance of the package on JWST/NIRCAM data of the Local Group dwarf irregular galaxy WLM $\left(M_v\approx-14.2 \ \text{mag}\right)$, which exhibits a complex, well-sampled CMD, and HST/ACS data of the ultra-faint Milky Way satellite dwarf galaxy Horologium I $\left(M_v\approx-3.7 \ \text{mag}\right)$, which has a much simpler but sparser CMD.
\end{abstract}



\section{Introduction} \label{sec:intro}

One of the most important relations in galaxy evolution is that between the mass of a dark matter halo and the stellar mass of the galaxy it hosts. Over the last decade, development on the modelling front has resulted in a variety of models that are able to reproduce observed galaxy luminosity functions at low to intermediate redshifts \citep[$z\leq3$, e.g.,][]{Benson2012,Vogelsberger2013,Vogelsberger2014a,Vogelsberger2014b,Torrey2014,Schaye2015,Asquith2018,Pillepich2018a,Dave2019}. However, significant differences remain in the particulars of how galaxies build their stellar masses in these models. For example, models with different implementations of supernova feedback exhibit differing degrees of ``burstiness'' in their star formation histories (SFHs); in models with strong, impulsive feedback, star formation in dwarf galaxies can exhibit variations on timescales of 10--100 Myr and this feedback can even drive the formation of dark matter cores and radial age gradients \citep[e.g.,][]{Governato2012,Madau2014,ElBadry2016,Burger2019,Graus2019}. Another difference between the models is the efficiency of environmental quenching; i.e., how rapidly the star formation in galaxies accreted onto more massive hosts is quenched. Observational studies have shown that the classical dwarf satellites \citep[$10^5 \leq$ \mstar $\leq 10^7$ \msun,][]{McConnachie2012} of the Milky Way (MW) exhibit a wide range of quenching times \citep[2--10 Gyr ago; see, for example, figure 12 of][]{Weisz2014a}, but they also fell into the MW halo at different times. Theoretical studies accounting for variation in the infall times of the classical dwarfs indicate that they were likely quenched 0.5--2 Gyr after they fell into the halo of the MW \citep{Fillingham2015,Wetzel2015,Engler2023}.

To test these models we require high-fidelity measurements of the SFHs of real galaxies for comparison. The color-magnitude diagram (CMD) modelling technique, which can achieve time resolutions of $\Delta t \sim 100$ Myr in the SFH, provides the highest precision SFH measurements for this purpose \citep[e.g.,][]{Tolstoy1996,Dolphin2002,Gallart2005}. In CMD modelling, individual stars in a galaxy are resolved in broadband imaging and their distribution in the CMD is modelled via the linear superposition of stellar population templates with different ages and metallicities. This technique can resolve stars across many phases of stellar evolution to precisely infer the SFH of the population, but typically requires deep photometry that reaches the oldest main sequence turn-off (MSTO) to constrain the earliest epochs of star formation \citep[e.g.,][]{Weisz2014a}. For more distant galaxies where the available imaging does not reach the MSTO, this method can still place constraints on recent star formation activity \citep[e.g.,][]{Bortolini2024a}. We will refer to a SFH measured with this method as a \emph{resolved SFH} throughout this paper to differentiate from other measurement techniques.

For example, there are other methods that are more suitable for measuring the SFHs of distant galaxies. Imaging surveys like the Sloan Digital Sky Survey \citep{Alam2015} and the Dark Energy Survey \citep{DESC2016} obtain multi-band photometric measurements for distant galaxies that sample the underlying spectral energy distributions (SEDs) of the galaxies, which can be used to obtain SFHs with relatively coarse time resolution ($\Delta t \sim 1$ Gyr). This is achieved by comparing the observed SEDs to large sets of template SEDs (often containing millions of different SFHs) constructed from stellar libraries \citep[e.g.,][]{Conroy2009,Conroy2010a,Conroy2010b,Pacifici2016,Leja2017,Baes2019a,Fioc2019,Smith2019,Suess2022,Wang2023}. Better time resolution can be achieved via full spectrum fitting, wherein an observed galaxy spectrum is modelled via the linear superposition of spectral templates for stellar populations with different ages, metallicities, and so on \citep[e.g.,][]{Cappellari2004,Koleva2009,Cappellari2017,Cappellari2023}. However, full spectrum fitting is still not as precise as CMD modelling because the light from all the stars in the population is mixed into a single spectrum rather than being resolved into individual stars.

The SFHs of low-mass dwarf galaxies are of special interest to several aspects of galaxy evolution and cosmology. As mentioned above, previous measurements of the resolved SFHs of the classical dwarf satellites of the Local Group have shown they have greatly reduced recent star formation rates (SFRs) compared to their cosmological averages \citep[e.g.,][]{Weisz2014a,Weisz2014b}. In concert with the fact that isolated dwarfs at these masses appear to be ubiquitously star-forming \citep{Geha2012,Phillips2014}, this suggests that environmental quenching processes are responsible for the recent quenching of the dwarf satellites. When combined with theoretical analyses that consider the possible infall histories of these satellites, the resolved SFHs can be used to constrain the timescales on which this quenching occurs \citep[e.g.,][]{Garling2024}. These environmental quenching timescales can then be compared to the results of semi-analytic models and hydrodynamic simulations to inform models of galaxy evolution. 

Additionally, the resolved SFHs of the ultra-faint dwarf (UFD) satellites of the Local Group (\mstar $\leq 10^5$ \msun) are of cosmological interest. Whereas the classical dwarf satellites of the Local Group appear to have only recently been quenched, the UFD satellites typically have ancient stellar populations consistent with having formed the majority of their stars ($\geq80\%$) prior to reionization \citep[$z_{re}=7.67\pm0.73$, $t_{re}=13.11\pm0.09$ Gyr,][]{Planck2020}, indicating that reionization likely played a significant part in their quenching processes \citep{Brown2014,Weisz2014a,Weisz2014b,Sacchi2021}. Significant uncertainty remains in the degree of spatial inhomogeneity (i.e., the ``patchiness'') of reionization, as dense regions of the Universe with many galaxies producing ionizing UV photons are expected to have been ionized earlier than underdense regions that host few galaxies. However, the timing of reionization as a function of environment depends on highly uncertain models of early galaxy formation, including the efficiency of star formation in extremely metal-poor gas and the UV luminosity function of Population III stars -- current models indicate variations in the timing of reionization of a few hundreds of Myr at the scale of the Local Group \citep[$r\approx5$--$10$ Mpc;][]{Ocvirk2020,Gnedin2022,Sorce2022,Balu2023}. Measuring the resolved SFHs of additional UFDs in the Local Group and comparing their quenching times may shed light on this important cosmological epoch. 
As the UFDs formed the majority of their stars so early, they can even help to constrain the nature of dark matter, as halo formation times can be altered under different dark matter models, which in turn affects how early gas can cool and stars can form in those halos \citep{Chau2017}.  

Surprisingly, there are no publicly-available modern implementations of the CMD modelling technique for measuring resolved SFHs, while several advanced implementations of SED fitting \citep[e.g., \textsc{prospector;}][]{Leja2017,Johnson2021,Suess2022} and full spectrum fitting \citep[e.g., \textsc{ppxf};][]{Cappellari2004,Cappellari2017,Cappellari2023} are publicly available. Given the importance of resolved SFH measurements as outlined above, we believe this represents an opportunity for innovation within the community. To this end, we develop and distribute the Julia package \textbf{StarFormationHistories.jl} for measuring resolved SFHs using the CMD modelling technique under the open-source MIT license.\footnote{The source code is hosted at \url{https://github.com/cgarling/StarFormationHistories.jl} and can be installed from the Julia general registry.} In \S \ref{sec:methods} we summarize the formalism of modern CMD modelling techniques and outline the methods we implement in our package. We illustrate the robustness of our methods when applied to idealized, synthetic data in \S \ref{sec:synthetic_data}. We then measure the SFH of the isolated WLM dwarf galaxy using JWST/NIRCAM data and compare to literature results in \S \ref{sec:wlm}. To illustrate the performance of our methods on sparser CMDs, we measure the resolved SFHs of the Horologium I UFD with HST/ACS data in \S \ref{sec:UFDs} and compare against literature results. In \S \ref{sec:discussion} we discuss opportunities for future development of our methodologies and software. 

\section{Methodologies in Resolved SFH Fitting} \label{sec:methods}

The CMD modelling technique achieves superior precision compared to SED fitting and full spectrum fitting because individual phases of stellar evolution can be resolved and compared to reference stellar evolution libraries in order to find the combination of stellar population properties that best reproduce the observed data. Consider a simple stellar population (SSP) of uniform age and metallicity. Stars with different initial masses will inhabit different areas of the CMD corresponding to unique phases of stellar evolution. Given sufficiently deep photometric data, a combination of multiple CMD features with age and metallicity dependence can be used to determine the properties of the SSP. Such sensitive CMD features include the MSTO, red clump, blue loop (i.e., the loop of red to blue helium burning stars), and red giant branch (RGB).

Such analyses are often leveraged to study star clusters, whose CMDs can typically be modelled with only one or two unique stellar populations \citep[e.g.,][]{VandenBerg2013,Perren2015,Leitinger2023}. Dwarf galaxies with \mstar $>10^5$ \msun (particularly those which are not satellites) have much more extended SFHs, which results in their CMDs being signficantly more complex. As noted by \cite{Dolphin1997}, the CMD of a composite stellar population is simply the superposition of the CMDs of its constituent populations.
This fact is fundamental to the CMD modelling techniques used to measure resolved SFHs.

However, there remain a number of choices which must be made regarding how to model observed CMDs. Stated simply, the stars we observe in a population can be viewed as a random realization drawn from an underlying continuous SFH that we seek to infer \citep[mathematically this can be described by a Poisson point process; see the discussion in][]{Gennaro2015}. We first must note that CMDs are continuous two-dimensional spaces and a choice must be made as to whether/how to discretize the space; this is discussed in \S \ref{subsec:discretization}. Next, a model must be developed that enables us to quantify the goodness-of-fit of a proposed SFH given the observed data and the available theoretical models (i.e., stellar evolution libraries). In a typical Bayesian analysis this would be a likelihood function of some kind, but likelihood-free inference methods may also be applied \citep[e.g.,][]{Gennaro2018}. As a complex population can be modelled as a superposition of many individual SSPs, we separate this into two steps. In \S \ref{subsec:templates} we formulate \emph{templates} that describe the expected distribution of stars in CMD-space for each SSP given the observational properties of the dataset (e.g., photometric uncertainty and completeness). The expected distribution of stars for a proposed complex SFH can then be constructed as a linear combination of these SSP templates where the weights on each template are SFRs. In \S \ref{subsec:binaries} we discuss how we construct templates that include unresolved photometric binaries. With the ability to construct these models, we discuss goodness-of-fit statistics in \S \ref{subsec:mdcomparison}, which we can use to derive maximum likelihood estimates for the SFRs using methods discussed in \S \ref{subsec:fitting} and sample the posterior of the SFRs as discussed in \S \ref{subsec:sampling}. Finally, the best-fit SFH for an observed population can depend strongly on how the population's stellar metallicity distribution is modelled. In \S \ref{subsec:hierarchical} we describe our formulation of hierarchical models in which we fit parametric age-metallicity relations simultaneously with the SFH in order to constrain the range of possible metal enrichment histories to a physically-motivated subset of possibilities.

\subsection{CMD Discretization} \label{subsec:discretization}
In SED fitting and full spectrum fitting the data are naturally discretized according to the photometric bandpasses and resolution elements of the spectrum, respectively, enabling straightforward application of statistics like $\chi^2$ to quantify the goodness-of-fit between models and data that share a discretization. This is not the case for the CMD, which requires either more complicated unbinned likelihoods or the imposition of an artificial discretization, usually in the the form of a binning scheme. While there is literature on unbinned methods for CMD modelling \citep[e.g.,][]{Naylor2006,Walmswell2013} and some interesting recent applications \citep[e.g.,][]{Gennaro2015,Gordon2016,Gennaro2018,Ramirez-Siordia2019,Chandra2021}, these methods are generally more computationally expensive and complex to implement than manual binning methods, particularly as they relate to deriving uncertainties on the best-fit SFHs \citep{Raja2003}. The vast majority of publications measuring resolved SFHs via CMD modelling use manual binning methods. Previous work has demonstrated that different choices in the binning strategy generally produce consistent results \citep[e.g.,][]{Monelli2010}.

The most common approach to manual binning of the CMD is to impose a regular grid along the two component axes and count the number of stars that fall into each bin, creating effectively a two-dimensional histogram. Such a 2-D discretization of the CMD is known as a \emph{Hess diagram}. With the CMD binned into a discrete Hess diagram, Poisson statistics can be used to quantify goodness-of-fit between the observed data and SSP models discretized to the same 2-D grid. This is the method used by the \textsc{match} software \citep{Dolphin2002,Dolphin2012,Dolphin2013}, one of the earliest and most prolific programs for measuring resolved SFHs \citep[interesting applications of \textsc{match} include][]{McQuinn2010,Weisz2011,Weisz2014a,Weisz2014b,Skillman2017,Hargis2020,Savino2023,McQuinn2024}. We likewise use Hess diagrams with regular grids as our means of CMD discretization, as use of a regular grid enables specialized algorithms for generating model templates for SSPs (described in detail in \S \ref{subsec:templates}).

An alternative approach involves discretizing the CMD space using an adaptive method that seeks to optimize some aspect of the discretization; an example would be a Voronoi tesselation designed to apportion equal numbers of stars to each bin such that the Poisson error per bin is roughly constant. Another is the approach adopted by the IAC-pop software, wherein the CMD is separated into regions capturing major evolution phases (e.g., the MS, RGB, etc.) and each region is discretized separately according to the stellar density in the region \citep{Aparicio2009}. This method can be beneficial as it allows for more uniform signal-to-noise across the discretization and presents an obvious way to mask areas of the CMD that should be excluded from the analysis (e.g., areas that contain foreground contamination in the observed data). This method additionally presents a straightforward way to study particular CMD regions; for example, the horizontal branch (HB) morphology remains quite difficult to model \citep[e.g.,][]{Savino2018} and it is common for the HB to be excluded from resolved SFH fits for this reason \citep[e.g.,][]{Savino2023}. However, these CMD regions generally have to be manually defined for every observed galaxy as the SFH and metallicity of the stellar population influence the CMD morphology in ways that cannot be determined automatically prior to fitting. As such, these region-based methods require more manual intervention than methods that use a uniform binning strategy. For this reason we do not implement a region-based discretization, but we provide methods for generating boolean masks to select specific portions of the Hess diagram for subsequent processing.

\subsection{SSP Template Construction} \label{subsec:templates}

Recall that a complex CMD containing multiple stellar populations is merely a superposition of the CMDs of the constituent populations \citep{Dolphin1997}. As a Hess diagram can be straightforwardly represented as a matrix, a superposition of the CMDs of several SSP becomes a sum of their Hess diagrams after discretization to a common grid. We will refer to the model Hess diagrams of SSPs as \emph{templates}, though they are often referred to as partial CMDs in other works. We prefer this nomenclature as our templates are always discretized and so are not technically CMDs.

Mathematically, a complex Hess diagram model can be constructed following equation 1 of \cite{Dolphin2002},

\begin{equation} \label{eq:composite}
  m_i = \sum_j r_j \, c_{i,j}
\end{equation}

\noindent where $m_i$ is the value of the complex model Hess diagram in bin $i$, $c_{i,j}$ is bin $i$ of template $j$, and $r_j$ is the multiplicative coefficient determining how significantly template $j$ contributes to the complex model. Defined in this way, the complex Hess diagram model $m$ is a linear combination of the templates $c_j$ with amplitudes $r_j$. It is typically suggested to normalize all templates to uniform SFR, for example, 1 \msun yr$^{-1}$. In this case the $r_j$ coefficients can be straightforwardly interpreted as SFRs.

We now turn to the critically important topic of how the templates are constructed. While the most straightforward method is to apply random sampling methods (discussed in Appendix \ref{appendix:montecarlo}), these methods create templates that suffer from Poisson error. It is preferable to devise a methodology whereby the template bins $c_{i,j}$ contain the expected number of stars per unit $r_j$ free from sampling noise. We accomplish this by using artificial star tests to model the two-dimensional probability distribution of observed magnitude and color given the intrinsic magnitudes of isochrone stars\footnote{We choose to refer to discrete points in an isochrone, which are uniquely identified by their initial stellar masses, as discrete "stars" for simplicity, though this is somewhat imprecise as the underlying stellar models are computed with a lower resolution grid of initial masses and later interpolated at intermediary initial stellar masses to increase the point density in the isochrone. See \cite{Dotter2016} for further discussion.}. Integrating these distributions over the Hess diagram grid for each isochrone star, additionally weighted by an IMF-related quantity, results in smooth templates where the value of bin $i$ of template $j$ (written as $c_{i,j}$ in the above equation) is the expected value for the number of stars that would fall into that bin per unit $r_j$ for a single SSP.

Let there be $N$ stars in isochrone $j$ with intrinsic magnitudes $m_{\text{int},N}$. In the context of Equation \ref{eq:composite}, we can calculate bin $i$ of the Hess diagram template for isochrone $j$ as the sum over the $N$ stars of the integrals of their two-dimensional probability distributions, $P(x,y \, | \, m_\text{int})$, over the bin multiplied by their detection probabilities, \precover, and an IMF weight, $w_\text{IMF}$, that quantifies how commonly each star occurs in a sampled stellar population \citep[this is similar to the occupation probability in][see their section 3.1]{Harris2001}. We write this as

\begin{equation} \label{eq:templatepix}
  c_{i,j} = \sum_N P_{\text{recover},N} \ w_{\text{IMF},N} \, \iint_i P(x,y \, | \, m_{\text{int},N}) \, dx \, dy.
\end{equation}

\noindent Below, we will show that the detection probability, \precover, and the two-dimensional probability distribution, $P(x,y \, | \, m_\text{int})$, can be derived from artificial star tests, while the IMF weight, $w_{\text{IMF},N}$, is a function only of the assumed stellar IMF and the initial masses of the stars in the isochrone. The derivation in this section is formulated for the case of single star systems; the generalization to unresolved photometric binaries is discussed in \S \ref{subsec:binaries}. 

\subsubsection{Artificial Star Tests}

We start from the fact that all CMD modelling techniques require accounting for observational effects, the most obvious of which are photometric uncertainties and point source incompleteness. It has long been standard practice to quantify these through artificial star tests \citep[ASTs; see, for example,][and references therein]{Annunziatella2013}. In short, ASTs take catalogs of artificial stars generated from a realistic distribution in the color-magnitude space and inject them into the science images using the observed point spread function with noise added to reflect the detector properties. By then attempting to recover the injected artificial stars using the same photometric pipelines used on the original science images, investigators can quantify their observational uncertainties. In particular, as the intrinsic magnitudes of the artificial stars are known, ASTs allow investigators to quantify the probability of detecting a star given its magnitude \precover and the probability distribution of its observed magnitude given its intrinsic magnitude $P(m_\text{obs} \, | \, m_\text{int})$. For photometric pipelines that perform detection using information from multiple filters simultaneously, these probabilities will be functions of intrinsic colors as well, but for presentational clarity we will consider the single-band derivation.

We next observe that many studies measuring resolved SFHs via CMD modelling only include the high-completeness portion of the Hess diagram in their fits \citep[\precover $\geq 0.5$; e.g.,][]{McQuinn2024}. This is due to the colluding factors at fainter magnitudes of poorer completeness, higher photometric errors, and greater contamination from background galaxies due to increased difficulty differentiating stars from background galaxies. In the limit of high completeness, it has been demonstrated that the distribution of photometric errors in ASTs can be modelled as a Gaussian distribution with bias $\mu$ and standard deviation $\sigma$ being functions of the intrinsic magnitude \citep[see, e.g., figure 20 of][]{Milone2012}. We can write this as $P(m_\text{obs} \, | \, m_\text{int}) \thicksim G \left[ \mu \left( m_\text{int} \right) + m_\text{int}, \sigma \left( m_\text{int} \right) \right]$. The bias $\mu \left( m_\text{int} \right)$ and standard deviation $\sigma \left( m_\text{int} \right)$ can be measured straightforwardly from catalogs of ASTs.

With models for the one-dimensional photometric uncertainty distributions $P(m_\text{obs} \, | \, m_\text{int})$, we must now combine them to form a two-dimensional distribution in CMD space. In the case that the magnitude on the $y$ axis does not appear in the $x$ axis color, the uncertainty distributions of $x$ and $y$ are separable and the two-dimensional probability distribution $P(x,y \, | \, m_\text{int})$ can be modelled as a two-dimensional Gaussian with a diagonal covariance matrix. The integral over the Hess diagram bins is fully analytic in this case. In the case that the $y$ axis magnitude \emph{does} appear in the $x$ axis color, as occurs when only two filters have been observed, then the covariance between the axes must be modelled as well. In this case, only the inner integral can be solved analytically; we complete the outer integral via Gauss-Legendre quadrature. 

\begin{figure*}[ht!]
  \centering
  \includegraphics[width=\textwidth,page=1]{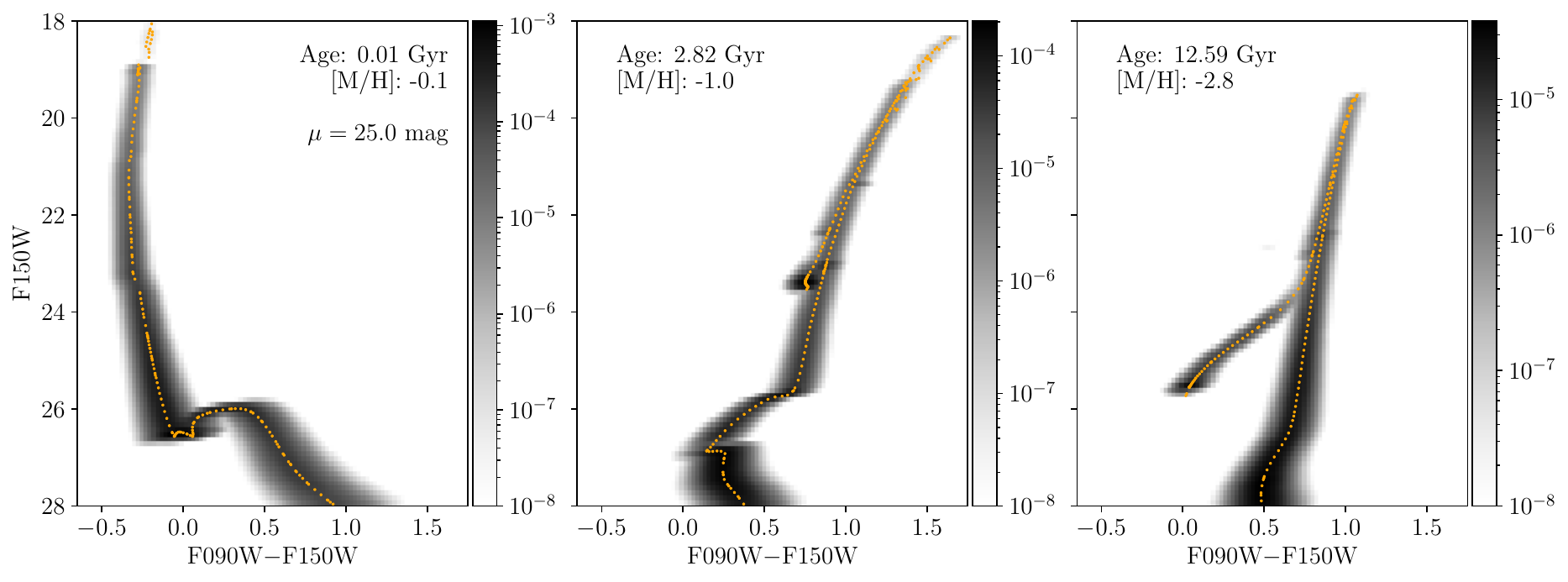}
  \caption{Hess diagram templates for three SSPs generated via the procedure outlined in \S \ref{subsec:templates}. We show templates for a young, metal-rich population (left), a population of intermediate age and metallicity (center), and an old, metal-poor population (right) to highlight how our method performs for populations with different CMD morphologies. The populations are modelled at a distance modulus $\mu=25$ mag ($d=1$ Mpc), consistent with that of the WLM dwarf irregular which we analyze in \S \ref{sec:wlm}. The photometric error models used to broaden the templates are based on our WLM analysis as well. The stars from the isochrones overplotted in orange. These templates are shown in units of expected number of stars per initial solar mass of stars formed in the population. These plots demonstrate that our method is capable of resolving a large dynamic range in expectation values while accurately tracking the isochrone through many phases of stellar evolution.}
  \label{fig:template_example}
\end{figure*}

\subsubsection{Stellar Initial Mass Functions}

The IMF weight $w_{\text{IMF},N}$ in Equation \ref{eq:templatepix} is the expected number of stars that would form within a small step in initial mass $\Delta M_\text{ini}$ per solar mass of stars formed; it therefore has natural units of $\text{M}_\odot^{-1}$. Let the initial masses of the stars in the isochrone, sorted from least to greatest, be $M$, and the probability distribution of initial masses for single stars given by the IMF be $dN/dM$, properly normalized such that it integrates to 1 over the range of possible initial stellar masses. The IMF weight for isochrone star $k$ is the number fraction of stars born between $M_{k}$ and $M_{k+1}$ divided by the mean mass per star born $\langle M \rangle$, such that the weight effectively represents the number of stars expected to be born with masses between $M_k$ and $M_{k+1}$ per solar mass of star formation.

\begin{equation} \label{eq:imfweights}
  \begin{aligned}
  w_{\text{IMF},k} &= \frac{ \int_0^{M_{k+1}} \frac{dN(M)}{dM} dM - \int_0^{M_{k}} \frac{dN(M)}{dM} dM }{\int_0^\infty M \times \frac{dN(M)}{dM} dM} \\
  &= \frac{ \int_{M_k}^{M_{k+1}} \frac{dN(M)}{dM} dM }{\langle M \rangle}.
  \end{aligned}
\end{equation}

\noindent We provide implementations of several popular IMFs \citep[e.g.,][]{Salpeter1955,Chabrier2001,Kroupa2001,Chabrier2003} in the companion package InitialMassFunctions.jl\footnote{The source code is hosted at \url{https://github.com/cgarling/InitialMassFunctions.jl} and can be installed from the Julia general registry.} that can be used to compute these weights. The general forms of these IMFs (e.g., power law, broken power law, lognormal) are exposed so that users can construct IMFs with different parameters as they wish. This design also enables IMF parameters (like a power law slope) to be optimized or marginalized over.

\subsubsection{Implementation}

Programmatically, we construct a matrix with all elements zero to represent an isochrone's Hess diagram, using the same bins as were used to discretized the observed CMD. For each star in the isochrone, we integrate its two-dimensional probability distribution $P(x,y \, | \, m_\text{int})$ over each bin that falls within $\pm5\sigma$ of the star's intrinsic center (truncated for efficiency) and add these values into their respective bins, additionally weighted by the star's detection probability and IMF weight.
By doing this for every star in the isochrone, we satisfy Equation \ref{eq:templatepix}, as each bin now contains the expected number of stars that should fall into that bin per solar mass of star formation. This normalization can be easily changed afterward if desired.

By explicitly modelling the two-dimensional probability distribution of the input stars from the isochrone and integrating them over the pixel-space of the Hess diagram, we are able to generate model Hess diagrams with precision that is independent of the bin sizes of the Hess diagram. This addresses the most significant challenge of the more traditional model Hess diagram construction methods that rely on Monte Carlo sampling.

Additionally, templates can be generated with this method in a few milliseconds each; in comparison, sampling $10^7$ stars to construct a template via the Monte Carlo method takes about a second using the methods outlined in Appendix \ref{appendix:montecarlo}. This has allowed us to perform all the analysis for this work on personal computers. Fitting resolved SFHs with other codes typically requires cluster-scale computing resources, which increases cost and complexity and can be a roadblock for junior researchers and those at institutions that lack such resources. Of course, the computational costs associated with performing the photometry and ASTs can still be substantial (for example, see section 3.7 and table 4 of \citealt{Weisz2024} for discussion of the ASTs performed on the JWST/NIRCAM imaging that we make use of in \S \ref{sec:wlm}). 

There is a practical matter to be considered of the spacing of stars along the isochrone. In order for our discretized method to be robust, the stars along the isochrone must be sufficiently dense in the CMD to produce a smooth final template free from discontinuities after discretization and smoothing by their respective probability distributions.
We therefore interpolate the stars in the isochrone such that the maximum spacing between adjacent stars is always less than the bin widths used to discretize the Hess diagram. This ensures there is sufficient overlap between adjacent isochrone stars to construct smooth templates without unnecessarily increasing the computational cost. Of course, a uniform spacing in magnitude does not correspond to a uniform spacing in the stellar initial masses, but the irregular sampling is dealt with by our IMF weighting, described above.

\begin{figure*}
  \centering
  \includegraphics[width=\textwidth,page=1]{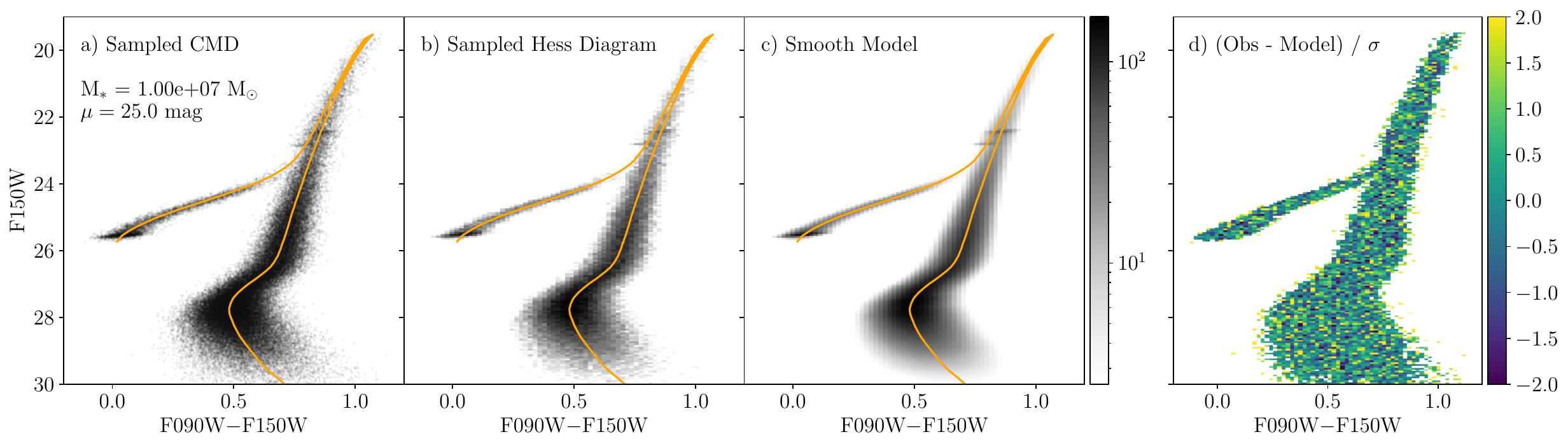}
  \caption{Comparison between a Hess diagram template constructed via random sampling and a smooth template generated with our method from \S \ref{subsec:templates}. Both methods use a PARSEC isochrone of age 12.6 Gyr with initial [M/H] $= -2.8$. The isochrone is overplotted in orange. The distance modulus and photometric error and completeness functions emulate those used in our analysis of the JWST/NIRCAM observations of WLM in \S \ref{sec:wlm}. \emph{a}) CMD sampled with the methods of Appendix \ref{appendix:montecarlo}. \emph{b}) Binned Hess diagram computed from the CMD. \emph{c}) Smooth template generated via our procedure described in \S \ref{subsec:templates}. \emph{d}) Residual between \emph{b} and \emph{c} in units of standard deviations (i.e., the residual significance).}
  \label{figure:template_compare}
\end{figure*}

\begin{figure}
  \centering
  \includegraphics[width=0.35\textwidth,page=1]{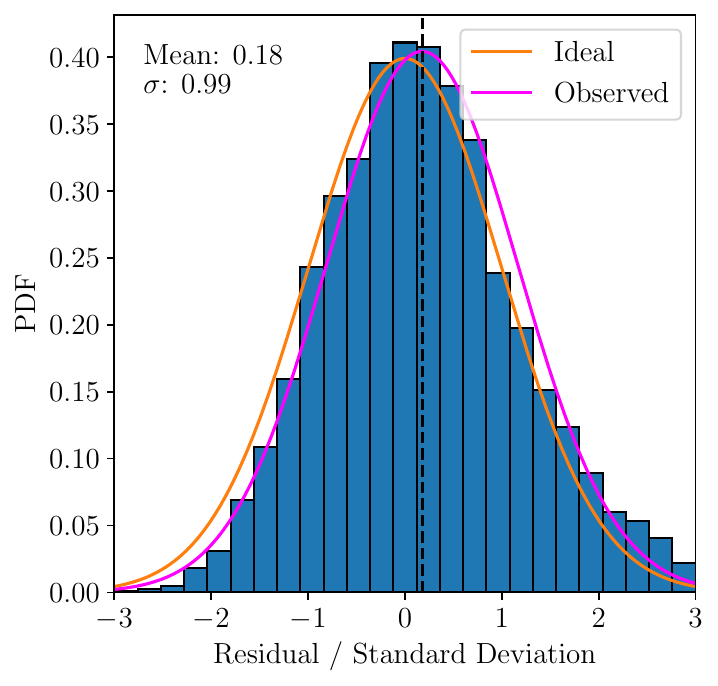}
  \caption{Distribution of the residual significance values from panel \emph{d}) of Figure \ref{figure:template_compare}. If our smooth template was a perfect model, the residual significance values would be normally distributed with mean 0 and standard deviation 1 (shown by the orange line). We achieve a standard deviation of $\sigma\sim1$, indicating that our model is robust. Our mean is slightly larger as we have excluded bins that are empty in the ``observed'' Hess diagram for presentational clarity. A normal distribution with these properties is shown in magenta for comparison.}
  \label{figure:sigma_dist}
\end{figure}

\subsubsection{Examples}

We show an example of three templates generated from PARSEC isochrones \citep{Bressan2012,Chen2014,Tang2014,Chen2015,Marigo2017,Chen2019a,Pastorelli2019,Pastorelli2020} with this procedure in Figure \ref{fig:template_example}. These isochrones use scaled-solar chemical compositions. The three isochrones chosen cover the range of stellar ages (0.01, 2.82, and 12.59 Gyr) and metallicities ([M/H] = -0.1, -1, and -2.8) relevant in the study of dwarf galaxy stellar populations. The populations are modelled at a distance modulus $\mu=25$ mag ($d=1$ Mpc) to match that of the WLM dwarf irregular which we analyze in \S \ref{sec:wlm}. The photometric error models used to broaden the templates are based on our WLM analysis as well. The templates are shown in units of expected number of stars per initial solar mass of stars formed in the population.

Note that the templates are shown with expectation values on a logarithmic scale, illustrating that our method is able to model the expectation values for the Hess diagrams of these SSPs with very high dynamic range. Also notable is the ability of our method to maintain its precision across many phases of stellar evolution, including very short-lived phases where template construction via random sampling suffers. Examples of this include the complicated morphology of the transition from the pre-main-sequence to the MS at apparent magnitude $\text{F150W}\approx26$ (absolute magnitude $\approx1$ mag) in the left panel and the short-lived thermally-pulsating asymptotic giant branch captured at apparent magnitude $\text{F150W}<19$ (absolute $\text{magnitude}<-6$ mag) in the center panel.


\subsubsection{Statistical Comparison to Random Sampling}

The accuracy of our template creation process can be further demonstrated by comparing against the random sampling method. Here we provide a quantitative comparison of a smooth Hess diagram template constructed with our method to a Hess diagram template constructed with the random sampling method described in Appendix \ref{appendix:montecarlo}. The code to reproduce this work is available from our source code repository.

We consider a PARSEC isochrone of age 12.6 Gyr with initial [M/H] $= -2.8$. For the observational model, we generally emulate the properties we assume for the JWST/NIRCAM data of WLM in \S \ref{sec:wlm}. We take a distance modulus of 25 mag and use simplified photometric error and completeness models that emulate those measured from the ASTs of the JWST/NIRCAM data presented in \cite{Weisz2024} and used to measure WLM's SFH with \textsc{match} in \cite{McQuinn2024}. For the template constructed via random sampling, we sample a total initial stellar mass (i.e., before losses due to stellar evolution) of $10^7$ M$_\odot$ such that we sample over ten million stars. We use this same stellar mass normalization when constructing the smooth template.

The results of this experiment are shown in Figure \ref{figure:template_compare}. The CMD of the randomly sampled population is shown in panel \emph{a}), which exhibits good sampling along the MSTO, but the sampling is worse near the TRGB as the interval of initial stellar masses that sample the TRGB is small. The CMD sampling is also poor at faint magnitudes where the point-source completeness drops rapidly. The corresponding Hess diagram is shown in panel \emph{b}), where we have chosen to construct 75 bins along the $x$ axis and 200 along the $y$ dimension. Our smooth template, constructed with the same isochrone and observational model, is shown in panel \emph{c}).

In a randomly sampled template such as that shown in panel \emph{b}), Poisson noise can be a significant source of random error. Statistically, each bin can be viewed as being sampled from a Poisson distribution with expected value $\lambda$. By definition, the Poisson distribution has variance equal to its expected value, so $\sigma^2=\lambda$. In the random sampling case, $\lambda$ increases linearly with the amount of stellar mass sampled. Clearly, as $\lambda$ increases, the signal-to-noise ratio per bin also increases as $S/N = \sqrt{\lambda}$. In contrast, our smooth model is constructed to provide a direct estimate of $\lambda$ with no random error. This enables us to perform a straightforward statistical comparison between the two Hess diagrams. In particular, the deviation of a Poisson variate $O$ from the expectation value $\lambda$ in units of the standard deviation is $S = \left(O - \lambda \right) / \sqrt{\lambda}$, which is often called the \emph{residual significance} or the Pearson residual. In the case that $\lambda$ is known perfectly and there are no other error sources, the distribution of $S$ should be Gaussian with mean 0 and standard deviation 1.

Panel \emph{d}) of Figure \ref{figure:template_compare} shows the residual significance for this example. In particular, we assume the bins of our smooth template as the expectation values $\lambda_i$ and compute the residual significances as $S_i = \left( O_i - \lambda_i \right) / \sqrt{\lambda_i}$, with $O_i$ being bin $i$ of the sampled Hess diagram shown in panel \emph{b}). Empty bins where $O_i=0$ have been made white for presentational clarity. Agreement between the observed Hess diagram and our model is excellent across the entire space. 

Note that the residuals in bins far from the isochrone sample the tails of the photometric error distribution and will always have positive residual significance $S_i$ in this plot. This is simply a Poisson sampling effect, as these bins have expectation values between 0 and 1 in the model, but a star was randomly sampled there in panel \emph{a}), leading to a positive value of $S_i$. These bins are offset by many more bins with expectation values between 0 and 1 that contain no observed stars in panel \emph{a}). All such bins with no observed stars are white in panel \emph{d}).

In our definition of $S_i$, we assumed that the smooth template contains expectation values $\lambda$ from which the Hess diagram bins $O_i$ of the randomly sampled populations were drawn. We can test this by looking at the distribution of $S_i$. For an ideal case where the underlying expectation values $\lambda$ are known perfectly and the only error source is the Poisson noise, $S_i$ should be normally distributed with mean 0 and standard deviation 1. We show our measured distribution in Figure \ref{figure:sigma_dist}. For clarity, we exclude all empty bins with $O_i=0$ from the distribution. As a result, the mean of the residual distribution will always be slightly greater than 0; the mean is consistent with 0 when these empty bins are included. Our method achieves a standard deviation of $\sigma\approx1$, indicating that our method is accurately modelling the distribution of the stars in the Hess diagram space, including the covariance between the \emph{x} and \emph{y} axes. Additionally, as our method integrates the intrinsic 2-D probability distributions to derive the occupation fraction in each Hess diagram bin (Equation \ref{eq:templatepix}), the precision of our template generation method is independent of the resolution of the Hess diagram grid, unlike most other algorithms.


\subsection{Unresolved Photometric Binaries} \label{subsec:binaries}

\begin{figure*}
  \centering
  \includegraphics[width=\textwidth,page=1]{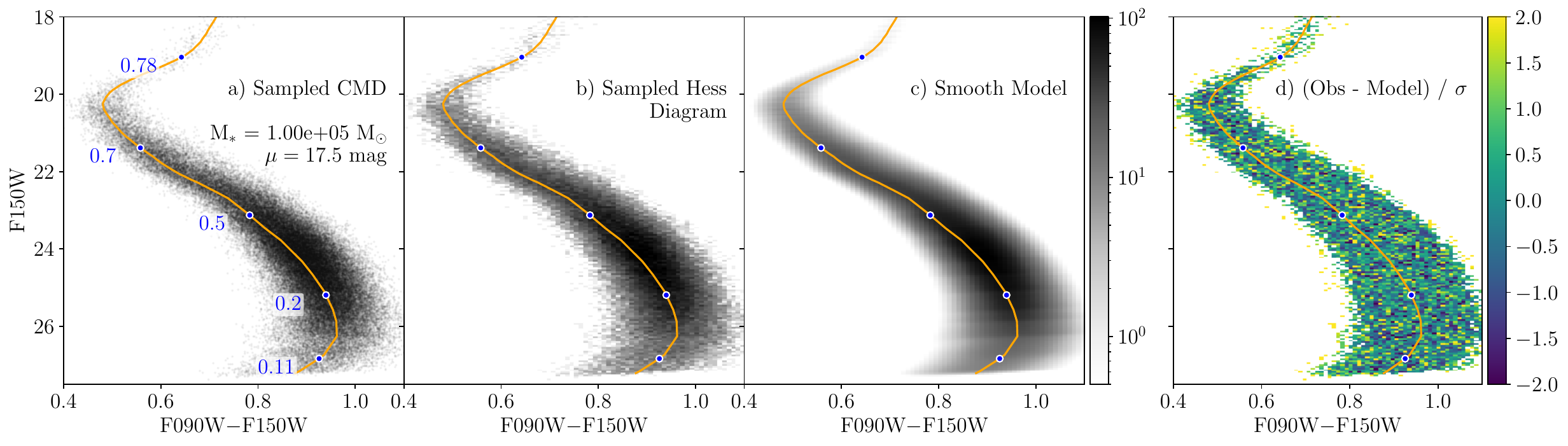}
  \caption{Analog of Figure \ref{figure:template_compare} computed with a binary fraction of $70\%$ where binary members are drawn from the IMF for single stars (i.e., independent draws). The same 12.6 Gyr, $\text{[M/H]}=-2.8$ PARSEC isochrone (orange line) and photometric uncertainty and completeness models are used, but the population is modelled at a distance modulus of 17.5 mag ($d\approx30$ kpc). Blue points labeled in panel \emph{a}) and overplotted in the other panels indicate stellar masses along the isochrone. The characteristic redward spread of the MS due to the high binary fraction is clearly visible and robustly modelled by the smooth template shown in panel \emph{c}).}
  \label{figure:template_compare_binaries}
\end{figure*}

The above discussion has presented our template construction procedure for single star systems. The primary effect of unresolved photometric binaries on the CMD is to broaden the lower MS and increase its spread towards redder colors \citep[e.g.,][]{Kroupa1991,Milone2012,Geha2013,DalTio2021}. While the observations of WLM analyzed in \S \ref{sec:wlm} do not reach sufficient depth for binaries to be significant, the observations of the UFD Horologium I analyzed in \S \ref{sec:UFDs} reach several magnitudes below the MSTO, suggesting that binaries may be important. More generally, JWST/NIRCAM and HST/ACS observations of UFDs at distances $<80$ kpc can be deep and precise enough to resolve MS broadening due to significant binary fractions \citep[$f_b \approx 0.4$; e.g.,][]{Geha2013,Spencer2018}. It is therefore worthwhile to consider how unresolved photometric binaries can be included in our method.

Single star systems and binary systems can be viewed as two subpopulations within a single SSP. As such, a composite template with a binary fraction $0 \leq f_b \leq 1$ is the weighted sum of the single-star template (binary fraction $0\%$) and a template representing only binaries (binary fraction $100\%$), with weights equal to the present-day population mass fractions in each subpopulation. In fact, computing the population mass fractions is the only step in the template creation process that is dependent on the binary fraction. As such, different binary fractions can be examined without having to recalculate the single-star or binary-only templates. This makes it exceedingly simple to fit the binary fraction, at least numerically -- a robust analysis also depends on very accurate modelling of the photometric uncertainties so the degree of MS broadening due to unresolved photometric binaries can be determined. In any case, we require a viable method to compute the binary template.

Unfortunately, the method we used for single star systems above does not generalize well to binary systems. As we integrate the probability distribution for each isochrone star individually over the bins of the Hess diagram, we maintain full accuracy in the sub-pixel location of the isochrone star. The downside of this approach is that it has a runtime complexity that is linear with the number of stars to be added into the model Hess diagram. For single stars, this is trivial, as an isochrone (even after interpolation) will typically contain only a few hundred stars. However, creating a template for binary systems requires fully sampling the range of both primary and secondary masses to avoid discontinuities in the model Hess diagram. Therefore the effective number of binary systems required is very high; if there are $N$ isochrone points in the single-star case, a naive implementation that constructs a binary system out of every pair of stars would produce $N \times (N-1) \div 2 \propto N^2$ systems which, using our method, would each be integrated across the Hess diagram pixel grid individually. The sub-pixel accuracy provided by our method is not worth this degree of computational cost. 

Instead, we use the early discretization technique as applied in \cite{Dolphin2002} and \cite{Bortolini2024}. Rather than integrating each star into the pixel grid individually (which maintains the sub-pixel positions of the stars exactly), isochrone stars are discretized onto the Hess diagram first. For all binary systems, we determine which Hess diagram bin the system falls into and add all of the system's probability weight into that single bin to construct a ``pure'' model Hess diagram; examples of these are shown in figure 3 of \cite{Dolphin2002} and figure A2 of \citealt{Bortolini2024}. After all isochrone stars have been sorted into bins, the Hess diagram is convolved with the photometric error and completeness functions. This approaches loses information on the sub-pixel positions of the isochrone stars, but removes the runtime scaling on the number of isochrone stars (or, in this case, binary systems) as the convolution runtime scales only with the size of the Hess diagram and the size of the convolutional kernel. 

Assuming one can compute a list of stellar binary pairs that adequately samples the range of primary and secondary masses, the remaining problem is to compute the analog of the IMF weight $w_\text{IMF}$ given in Equation \ref{eq:imfweights} for each binary system. Let $M_p$ be the sorted list of initial masses for primary stars and $M_s$ be the sorted list of initial masses for secondary stars. Conceptually, the IMF weight for a binary system with primary mass $M_{p,i}$ and secondary mass $M_{s,j}$ is the number fraction of binary systems born with primary masses between $M_{p,i}$ and $M_{p,i+1}$ and secondary masses between $M_{s,j}$ and $M_{s,j+1}$ per unit solar mass formed.

Clearly the IMF weight depends on the primary and secondary stellar initial mass functions. The simplest case is that of ``independent draws,'' where binary pairs are simply formed out of random, independent draws from the initial mass function for single stars. While newer analyses prefer ``correlated draws'' which sample from a stellar system mass function and then apportion mass to primaries and secondaries following a binary mass ratio distribution \citep[e.g.,][]{Goodwin2013}, studies of dwarf galaxies have shown that independent draws can adequately model these objects in most cases \citep[e.g.,][]{Geha2013, Gennaro2018}. We implement both methods in our Monte Carlo sampling methods (see Appendix \ref{appendix:montecarlo}), but we implement only independent draws in our smooth template modelling procedure in our initial release.

Figure \ref{figure:template_compare_binaries} illustrates the performance of our binary template construction algorithm. This figure is analogous to Figure \ref{figure:template_compare}, using the same isochrone and photometric uncertainty and completeness models, but the population with stellar mass $10^5$ M$_\odot$ is modelled at $d\approx30$ kpc with a binary fraction of $70\%$ assuming independent draws. The redward spread of the MS characteristic of high binary fractions is clearly visible and our smooth template shown in panel \emph{c}) is in excellent agreement with the Monte Carlo sampled population. The residual significance distribution of this example is of equal quality to that shown for the single-star case in Figure \ref{figure:sigma_dist}, achieving a standard deviation $\sigma\approx1$. The code to reproduce this example is available from our source code repository.

\subsection{Statistical Model -- Data Comparisons} \label{subsec:mdcomparison}
Once the SSP templates have been generated, we come to the topic of how to quantify the goodness-of-fit for a proposed set of coefficients $r_j$ as defined in Equation \ref{eq:composite}. A thorough discussion of this topic is presented in section 2.3 of \cite{Dolphin2002}; we will summarize some of their key conclusions here. As an observed Hess diagram is essentially a 2-D histogram, each bin in the observed Hess diagram $n_i$ can be thought of as a Poisson variate, and we are modelling the expectation value of that variate as $m_i$ from Equation \ref{eq:composite}. As such, it is logical to use a goodness-of-fit statistic derived, in some way, from the Poisson likelihood. The Poisson likelihood of observing $n_i$ stars in bin $i$ given an expectation value of $m_i$ is

\begin{equation} \label{eq:poissonlikelihood}
  P_i = \frac{m_i^{n_i}}{\exp{\left(m_i\right)} \, n_i!}.
\end{equation}

\noindent \cite{Dolphin2002} argue for the use of the Poisson likelihood ratio (PLR), defined as the ratio of the probability of drawing $n_i$ points from a Poisson distribution with expectation value $\lambda=m_i$ to that of drawing $n_i$ points with $\lambda=n_i$. For a single bin $i$, this is 

\begin{equation} \label{eq:plr}
  PLR_i = \frac{m_i^{n_i} \, \exp{\left(n_i\right)}}{n_i^{n_i} \, \exp{\left(m_i\right)}}
\end{equation}

\noindent which no longer involves the $n_i!$ term in the denominator of Equation \ref{eq:poissonlikelihood}. Taking the product over all bins gives the cumulative likelihood ratio of the proposed model. For the purposes of optimization and sampling, the negative logarithm of the cumulative PLR is more useful:

\begin{equation} \label{eq:logplr}
  -\text{ln}\left(PLR\right) = \sum_i m_i - n_i + n_i \, \text{ln} \left( \frac{n_i}{m_i} \right).
\end{equation}

\noindent This statistic has a number of useful properties. \cite{Dolphin2002} show that this statistic has only one minimum with respect to the fitting coefficients $r_j$, enabling use of any minimizer that is convenient. The gradient with respect to the fitting coefficients is also analytic and simple to calculate, enabling the use of gradient-accelerated optimizers and sampling methods. Due to these properties, and the historic success of this statistic as implemented in \textsc{match} \citep{Dolphin2002} and other works \citep[e.g.,][]{Geha2013}, we implement the PLR as our fitting statistic. When discussing our objective function, we are referring to Equation \ref{eq:logplr}.

\subsection{Fitting Model Parameters} \label{subsec:fitting}
There are many numerical algorithms appropriate for optimizing the SFR coefficients. For example, \textsc{sfera} \citep{Cignoni2015,Bortolini2024} and IAC-pop \citep{Aparicio2009} use genetic algorithms for their strong global convergence properties. In our experience, modern algorithms that utilize gradient information (i.e., first-order and quasi-Newton methods) can fit the $r_j$ of simple models described by Equation \ref{eq:composite} robustly with excellent performance. One could even use second-order methods as the Hessian of the objective (Equation \ref{eq:logplr}) is analytic and positive-definite \citep[see section 2 of][]{Walmswell2013}. However, the Hessian of the objective is dense and so the increase in convergence efficiency from second-order methods does not make up for the computational cost of evaluating the Hessian. We find success with the Broyden–Fletcher–Goldfarb–Shanno \citep[BFGS; see section 6.1 of][]{Nocedal2006} family of minimization algorithms, which require only the objective and its gradient with respect to the fitting parameters. These are quasi-Newton methods that internally construct an approximation to the Hessian (or, more conveniently, its inverse), giving them convergence properties rivaling second-order methods without incurring the computational cost of computing the Hessian directly.

The only constraint to be considered is $r_j\geq0$, as negative star formation rates would be unphysical. Such a constraint can be easily dealt with by the L-BFGS-B algorithm \citep{Zhu1997}. Implemented originally in Fortran, this method is a variant of the limited-memory BFGS algorithm (L-BFGS) that supports simple upper and lower bounds on the fitting variables. In the original BFGS method, the Hessian approximation is a dense matrix of dimensionality $(n,n)$ when fitting $n$ parameters. In the L-BFGS method, the Hessian is approximated with a limited number of vectors, which can save significant memory for large $n$. The Fortran implementation of L-BFGS-B is now available via wrappers in many higher level languages; for example, it is well-known as a backend for SciPy's minimization routines. We provide a method to optimize the coefficients $r_j$ in Equation \ref{eq:composite} with L-BFGS-B via the \href{https://github.com/Gnimuc/LBFGSB.jl}{LBFGSB.jl} wrapper package. For a Hess diagram with 100 bins in each dimension, this method can solve for the $r_j$ of 200 templates in 15 ms; constructing 200 templates might take 1.5s, so the solution time is negligible in comparison. 

While the above approach is exceedingly simple and quite robust for simple cases, the more general solution to the problem is to perform a transformation of variables so that we can utilize optimizers (and later, samplers) that require the fitting variables to be continuous and unconstrained. As we are using the PLR as our fitting statistic, our optimization is a form of maximum likelihood estimation. Such maximum likelihood estimates (MLEs) are invariant to variable transformations. So long as we are interested only in the MLE, we may fit the $r_j$ under any convenient transformation that maps all real numbers to the positive reals. This topic is discussed by \cite{Dolphin2013} in the context of sampling methods, which we will cover later. They suggest fitting variables $a_j$ such that $r_j=a_j^2$ or $r_j=|a_j|$ which satisfy the constraint $r_j\geq0$ for all real $a_j$. Both of these transformations result in objectives that are symmetric about $a_j=0$; i.e., the objective is the same at $a_j=-1$ as it is at $a_j=1$ because these values map to the same $r_j$. We prefer the parameterization $r_j=\exp{\left(a_j\right)}$ as the constraint $r_j\geq0$ can be achieved while maintaining a unique mapping between all $a_j$ and $r_j$. Under this transformation, we are free to fit the $a_j$ without constraints using any method we choose.

This exponential transformation requires marginally more BFGS iterations to achieve equivalent convergence criteria compared to $r_j=a_j^2$, but this minor increase in cost is offset by the fact that the exponential transformation allows us to obtain random uncertainty estimates on the $r_j$ from a BFGS optimization for free. We will present a short derivation here, but refer the reader to \cite{Dovi1991} and Appendix A of \cite{Yuen2010} for more information. Note that the Poisson distribution for large expectation value $\lambda$ is very similar in shape to a Gaussian distribution with mean $\lambda$ and variance $\lambda^2$ \citep[see, for example, equation 11.11 of][]{Taylor1997}. For a Gaussian loglikelihood objective function, it can be shown that the Hessian matrix is equal to the inverse of the Gaussian covariance matrix. Therefore the Hessian is constant for all parameters; since the loglikelihood is quadratic in the fitting variables, its second derivatives are all constants. As such, if one can calculate the Hessian from the Gaussian objective one can obtain the covariance matrix of the parameters in the fit. A similar calculation can be applied to other objectives to obtain \emph{approximations} of the covariance matrices of their parameters. The validity of these approximations depends on the shape of the objective function in question.

In general most objectives will not have constant Hessians. However, if an objective is roughly quadratic in the vicinity of the MLE, then the Hessian evaluated at the MLE is an estimator for the covariance matrix of the parameters. In other words, if the objective in the vicinity of the MLE can be well-approximated as a Gaussian objective, then the Hessian can be used to estimate the covariance matrix of the fitting parameters. Once such a covariance matrix is obtained, one can obtain standard errors by taking the square root of its diagonal, or even draw samples from it to include parameter covariances. However, whenever such approximate error estimates are used one should always demonstrate they are consistent with more robust estimates that make no assumptions about the properties of the objective. In \S \ref{sec:synthetic_data} we provide examples that demonstrate that we can use the Hessian approximation generated during a BFGS optimization to derive useful uncertainty estimates on the fitted $r_j$ that are broadly consistent with results from much more expensive sampling methods.

We note that this method for quantifying random uncertainties should not be applied when the MLE contains many $r_j$ consistent with 0. As the Poisson likelihood diverges from the Gaussian likelihood as the expectation value $\lambda \to 0$, the approximation that our PLR objective is quadratic in the vicinity of the MLE likewise breaks down, resulting in systematically underestimated random uncertainties. 

\subsection{Sampling Model Parameters} \label{subsec:sampling}
More robust random uncertainties on the fitting parameters can be derived if the objective can be measured or sampled without making assumptions about its statistical properties. Sampling methods as applied to resolved SFHs are considered by \cite{Dolphin2013} and we encourage interested readers to review their work. Our experience has supported their conclusions. We summarize some of these conclusions and discuss a few ways in which our sampling methods differ from those proposed by \cite{Dolphin2013} below.

It is now common practice in astronomy to use Markov Chain Monte Carlo methods (MCMC) to draw samples from distributions of interest. The affine-invariant ensember sampler, popularized by the \textsc{emcee} implementation \citep{Foreman-Mackey2013}, is a particularly widespread variant. This method is convenient as it requires only evaluations of the target distribution (\emph{not} its gradient) and can be very computationally efficient if the target distribution is cheap to evaluate. However, its performance can suffer in large dimensional problems. For this reason, \cite{Dolphin2013} find that sampling methods like these are inefficient for estimating the random uncertainties of the fitting coefficients $r_j$. For typical fits with more than 100 templates we likewise find these types of MCMC methods to be inefficient. However, for smaller numbers of templates (as one might use when fitting globular clusters), these methods can be competitive. For this reason we provide a method that uses the affine-invariant ensember sampler to provide samples of the fitting coefficients $r_j$. We have not used this sampling method in this work but we have verified that it is robust for small numbers of templates ($\sim20$).

For large dimensional problems, greater sampling efficiency can be achieved with other MCMC variants. \cite{Dolphin2013} utilize the Hybrid Monte Carlo algorithm developed by \cite{Duane1987} to efficiently sample the high-dimensional parameter space. This method has proven to be of great use in a variety of fields over the years, leading to innovations in algorithm design that have made implementations increasingly robust and efficient. The algorithm is now more commonly referred to as Hamiltonian Monte Carlo (HMC), due to the foundations of the algorithm being Hamiltonian dynamics. A comprehensive 59-page review of the theoretical formulation of the algorithm is given in \cite{Betancourt2017} and is outside the scope of this work. We will comment briefly that a Markov transition in HMC from position $p_1$ to position $p_2$ can be understood as the integration of a trajectory obeying Hamilton's equations through the parameter space that begins at $p_1$ and ends at $p_2$ with the objective serving as the potential energy distribution. The acceptance probability is the ratio of the final Hamiltonian to that at the beginning of the trajectory, which approaches 1 as the numerical integration error approaches 0 and the Hamiltonian is fully conserved. This fact reveals the usefulness of the HMC method. As the number of model parameters increase, the acceptance probabilities for Markov transitions in MCMC variants like the affine-invariant ensember sampler constantly decrease. HMC takes longer to generate each Markov transition because it needs to perform a numerical integration, but the probability of those transitions being accepted remains extremely high even for high-dimensional objectives.  

However, there are several algorithmic hyperparameters that can significantly affect the performance and accuracy of the algorithm. The gradient of the objective is used to inform the initial direction of the trajectories, but the kinetic energies, integration step sizes, and integration lengths of the trajectories are all free parameters. A poor choice of kinetic energies and integration lengths can lead to Markov transitions that are very short and do not effectively explore the objective. For optimal sampling efficiency the integration step size should be as large possible while adequately conserving the Hamiltonian. These hyperparameters can be tuned during warm-up procedures that probe the geometry of the objective. In particular, we use the No-U-Turn Sampler developed by \cite{Hoffman2014} as implemented in the DynamicHMC.jl \citep{DynamicHMC.jl} package. This is the principle method we use to estimate random uncertainties on model parameters in this work.

As we pointed out in the last section, MLEs are invariant under variable transformations, making them trivial to implement. Sampling methods, however, are not. This is a consequence of MLEs being point estimates while sampling methods trace \emph{volumes}, and volume elements are not conserved under changes of variables. This effect can be accounted for by multiplying the transformed distribution by the magnitude of the Jacobian determinant of the transformation \citep[see, e.g., section 2.12 of][]{Bilodeau1999}. If $\mathbf{r}$ is the set of all $r_j$, $f(\mathbf{r})$ is our original distribution, and we wish to perform a change of variables to sample $\boldsymbol{\theta}=\ln{\left(\mathbf{r}\right)}$, it can be shown that in the continuous case 

\begin{equation}
  \begin{aligned}
    \int_A f(\mathbf{r}) \, d \mathbf{r} &= \int_{\ln{\left(A\right)}} f \left( \exp{\left(\boldsymbol{\theta}\right)} \right) \, | J \left(\boldsymbol{\theta} \to \mathbf{r} \right) | \, d \boldsymbol{\theta} \\
    | J \left(\boldsymbol{\theta} \to \mathbf{r} \right) | &= \mathbf{r} = \exp\left(\boldsymbol{\theta}\right) \\
    \int_A f(\mathbf{r}) \, d \mathbf{r} &= \int_{\ln{\left(A\right)}} \exp\left(\boldsymbol{\theta}\right) \, f \left( \exp{\left(\boldsymbol{\theta}\right)} \right) \, d \boldsymbol{\theta}. \\
  \end{aligned}
\end{equation}

\noindent As means and variances are integrals in the continuous case, it is clear that discrete samples must follow the distribution with the Jacobian correction in order for their statistical moments to be correct. We apply this Jacobian correction to our sampling methods that utilize change of variables, and additionally apply it to the Hessian-based random uncertainty estimates discussed in the previous section. As we do not apply any explicit priors on the model parameters, we implicitly assume that all $r_j \geq 0$ are equally likely. 

\subsection{Hierarchical Models} \label{subsec:hierarchical}
The basic formula given in Equation \ref{eq:composite} constructs a complex model Hess diagram as the linear combination of $j$ individual templates. These templates can be constructed to model the Hess diagrams of SSPs with any combination of age and metallicity for which isochrones are available. As such, this approach places no constraints on the metallicity distribution function (MDF) or age-metallicity relation (AMR; $\langle[\text{M}/\text{H}]\rangle(t)$) of the population being modelled. This is in obvious disagreement with classical chemical enrichment models where stellar processes (e.g., supernovae explosions and AGB winds) create and distribute metals into the interstellar medium (ISM), leading to broadly increasing metallicities over time \citep[e.g.,][]{Lanfranchi2003,Lanfranchi2004}. Such freedom in the AMR can result in unphysical solutions where the AMR implied by the best-fit $r_j$ is highly variable and non-monotonic; in turn, unphysical AMRs generated by simply fitting the $r_j$ in Equation \ref{eq:composite} can significantly bias the recovered SFHs. It is therefore necessary to formulate more complex models that limit the range of possible AMRs to a more realistic subset. Several approaches to solving this problem have been explored in the literature. We will describe a few below for context, then describe the hierarchical models we implement. 

\textsc{match} supports a non-parametric AMR where $\langle[\text{M}/\text{H}]\rangle(t)$ is essentially unconstrained. It also implements simpler parametric forms that can be constrained to be monotonically increasing towards the present day; this is the approach we choose to take here. Although detailed chemical enrichment models indicate some Local Group dwarfs may have accreted small amounts of pristine gas at late times \citep[e.g.,][]{Kirby2013}, the assumption that the AMR increases monotonically to the present-day is generally acceptable for resolved SFH studies \citep[see, e.g., appendix B of][]{Savino2023}. At each time $t$, \textsc{match} uses only a single template with age $t$ and metallicity $\langle[\text{M}/\text{H}]\rangle(t)$ to represent the stars forming at that time; i.e., the only spread in metallicity at fixed $t$ is whatever spread was chosen by the user when generating the template, which is typically 0.1--0.2 dex \citep[e.g.,][]{Weisz2011}, such that the number of free metallicity-related parameters is equal to the number of age bins. Both types of models greatly reduce the parameter space of possible template combinations by restricting the allowed spread in metallicity at fixed time. The non-parametric form allows additional freedom for the AMR to respond to properties of the SFH; for example, if the best-fit SFH results in a SFR of 0 between times $t_1$ and $t_2$ then the change in metallicity between these times can be 0 as well. 

While the non-parametric AMR model of \textsc{match} provides a good mix of flexibility and robustness, there is no direct relation between $\langle[\text{M}/\text{H}]\rangle(t)$ and the fitted SFRs. Given that dwarf MDFs can generally be modelled with relatively simple singe-zone chemical evolution models \citep[e.g.,][]{Kirby2011,Kirby2013}, it is conceivable that one could use the fitted SFRs as input to such a model to determine a self-consistent AMR. IAC-star \citep{Aparicio2004} supports such analyses, though it requires some care in the choice of model parameters (e.g., the metal yield per solar mass of stars formed) to ensure the AMRs are realistic. For example, IAC-star supports both outflows and inflows in its chemical evolution model; if the parameters used to model these processes are poorly chosen, unphysical AMRs may still result. However, if these parameters are chosen well, this model is attractive as it can generate a fully self-consistent AMR and SFH simultaneously. This approach would be even more interesting if the parameters of the chemical evolution model could be fit simultaneously with the SFH. In the future we plan to explore this possibility, but presently we restrict our considerations to a few special cases. 

We implement two parametric AMRs in our initial release. The first, which we use to model WLM in \S \ref{sec:wlm}, is a linear model $\langle[\text{M}/\text{H}]\rangle(t)=\alpha \left( T_{max} - t \right) + \beta$ with $\alpha\geq0$ which can be produced in chemical models with exponentially increasing star formation rates. Here $T_{max}$ is the maximum lookback time for which the AMR is valid (at which time the mean metallicity is $\beta$) and $t<T_{max}$. The second is linear in the metal mass fraction $\langle Z \rangle (t) = \alpha \left( T_{max} - t \right) + \beta$ with $(\alpha,\beta) \geq 0$ such that $\langle[\text{M}/\text{H}]\rangle(t)$ is logarithmic with time. Such a chemical evolution can be produced in a closed box chemical model with constant star formation efficiency (see, e.g., section 12.2.1 of \href{https://galaxiesbook.org/chapters/II-05.-Chemical-Evolution.html\#The-closed-box-model}{\emph{Dynamics and Astrophysics of Galaxies}} by Jo Bovy, in preparation). These are special cases of the chemical evolution models supported in IAC-star \citep{Aparicio2004} that we will show are capable of producing robust SFHs.

We are now faced with a choice of how to construct our final model Hess diagram given our AMR. Both \textsc{match} and IAC-star make use of internal stellar evolution libraries and bolometric correction tables to interpolate isochrones at the exact age $t$ and mean metallicity $\langle[\text{M}/\text{H}]\rangle(t)$ they require. \textsc{match} always uses a single template for each unique time $t$ with a user-specified metallicity dispersion \citep[typically 0.1 -- 0.2 dex; e.g.,][]{Weisz2011} while IAC-star supports adding some user-defined metallicity dispersion for fixed $t$ but defaults to a single template as well.

Our solution is to model the MDF at fixed time $t$ as a Gaussian distribution with mean $\langle[\text{M}/\text{H}]\rangle(t)$ and a constant standard deviation $\sigma$ which is typically 0.1--0.2 dex but can be fit simultaneously with the SFH and AMR. This approach is motivated by recent simulation results that show metal mixing in dwarfs occurs on timescales 0.1--1 Gyr \citep{Emerick2020}, contrary to the often-adopted approximation of instantaneous mixing which implies all stars born at the same time should have the same metallicity. Long metal mixing timescales imply significant gas-phase metallicity dispersion which could create dispersion in the metallicities of stars born in different regions of the galaxy. 

This approach makes the most sense in the context of an isochrone grid, defined by the outer product of $j$ isochrone ages $t_j$ and $k$ isochrone metallicities [M/H]$_k$ for a total of $j \times k$ isochrones. This is a common way for pre-calculated isochrones to be distributed. Given the available templates derived from such an isochrone grid, relative weights are assigned to each template following Gaussian distributions with means $\langle[\text{M}/\text{H}]\rangle(t_j)$ and a constant standard deviation $\sigma$, normalized such that the sum over $k$ for all relative weights with fixed $t_j$ is 1. This is a hierarchical model as it enables us to still use Equation \ref{eq:composite} to construct our composite Hess diagram, but we do not have to fit the per-template coefficients, which we write as $r_{j,k}$ in the case of an isochrone grid. Instead we fit coefficients $R_j$ that are effectively SFRs for each unique isochrone age $t_j$ and distribute that SFR amongst all templates with age $t_j$ according to the adopted Gaussian MDF and isochrone metallicities [M/H]$_k$. Letting $\mu_j = \langle[\text{M}/\text{H}]\rangle(t_j)$, we can write this as

\begin{equation} \label{eq:hmodel}
  \begin{aligned}
    a_{j,k} &= \exp{ \left(-\left(\frac{[\text{M}/\text{H}]_k - \mu_j}{\sigma}\right)^2\right)} \\
    r_{j,k} &= R_j \frac{a_{j,k}}{\sum_k a_{j,k}} \\
    m_i &= \sum_{j,k} r_{j,k} \, c_{i,j,k} \\
  \end{aligned}
\end{equation}

\noindent where the final expression is the extension of Equation \ref{eq:composite} to the case of an isochrone grid where templates are indexed by their age $t_j$ and metallicity [M/H]$_k$.

Such a hierarchical model has a number of attractive properties. We can continue to use the PLR as our fitting statistic and, as discussed in \S \ref{subsec:mdcomparison}, the gradient of the PLR with respect to the $r_{j,k}$ are analytic and efficient to calculate. Liberal application of the chain rule allows one to derive fully analytic expressions for the gradient of the PLR with respect to the $R_j$ as well as the AMR parameters $\alpha$, $\beta$, and $\sigma$ in such a hierarchical model, allowing us to utilize the same efficient and robust gradient-informed solvers and samplers as we used in the case of the original model. Derivations of these gradients are given in our online documentation. Other formulations require application of less efficient numerical methods. For example, IAC-pop \citep{Aparicio2009}, which performs fitting of templates generated from IAC-star \citep{Aparicio2004}, uses the genetic algorithm of \cite{Charbonneau1995} which cannot use gradient or curvature information. Another benefit to this approach is that, as the SFRs and AMR parameters are all optimized and sampled simultaneously, we can fully capture covariance between the metallicity-related parameters and the SFRs. 

Of course, a model's convenience is secondary to its ability to reproduce observed data. In this respect there are a few aspects of our model we wish to discuss. Our method requires the isochrone grid to be reasonably dense in metallicity; we see good results with a metallicity spacing of $\Delta[\text{M}/\text{H}]\approx0.05$--0.1 dex, as we typically find the best-fit Gaussian MDF width for stars born at a fixed time to be 0.1--0.2 dex. For metallicity spacings approaching or exceeding 0.25 dex, we see significantly degraded fits for most objects.

Additionally, the PLR fitting statistic is very senstive to changes in the AMR parameters $\alpha$, $\beta$, and $\sigma$, as changes to these parameters effect changes in the $\mu_j$ and therefore every coefficient $r_{j,k}$ in Equation \ref{eq:hmodel}. While the slope $\alpha$ and intercept $\beta$ typically converge robustly, we occasionly find issues with the Gaussian width $\sigma$ as it is weakly covariant with the photometric error model used to broaden the SSP templates. Consider the representative case of a UFD with a predominantly old, metal-poor population. If the photometric errors calculated from the ASTs are systematically underestimated (i.e., the true photometric errors in the data are larger than indicated by the ASTs), the generated SSP templates will have sharper features than the observed CMD. This can be somewhat offset by fitting a large $\sigma$ (e.g., $\sigma \geq 0.3$ dex) as a spread in metallicity can emulate spread due to photometric uncertainties. Therefore, it is often advisable to fix $\sigma$ in the range 0.1--0.2 dex rather than fitting it. \textsc{match} typically uses a similar metallicity dispersion when generating templates \citep[e.g.,][]{Weisz2011}.

Another potential criticism of the parametric AMR models used in this work is that these AMRs are always increasing, even when there is no active star formation. We plan to add other AMR models in the future to address this issue, but in our present work we find we are still able to robustly measure resolved SFHs even with this limitation. In \S \ref{sec:wlm}, we show that the non-parametric AMR fit with \textsc{match} for WLM \citep[figure 7 of][]{McQuinn2024} is very similar to the result derived by applying our linear [M/H] AMR model to the same dataset. 

\begin{figure*}
  \centering
  \includegraphics[width=\textwidth,page=1]{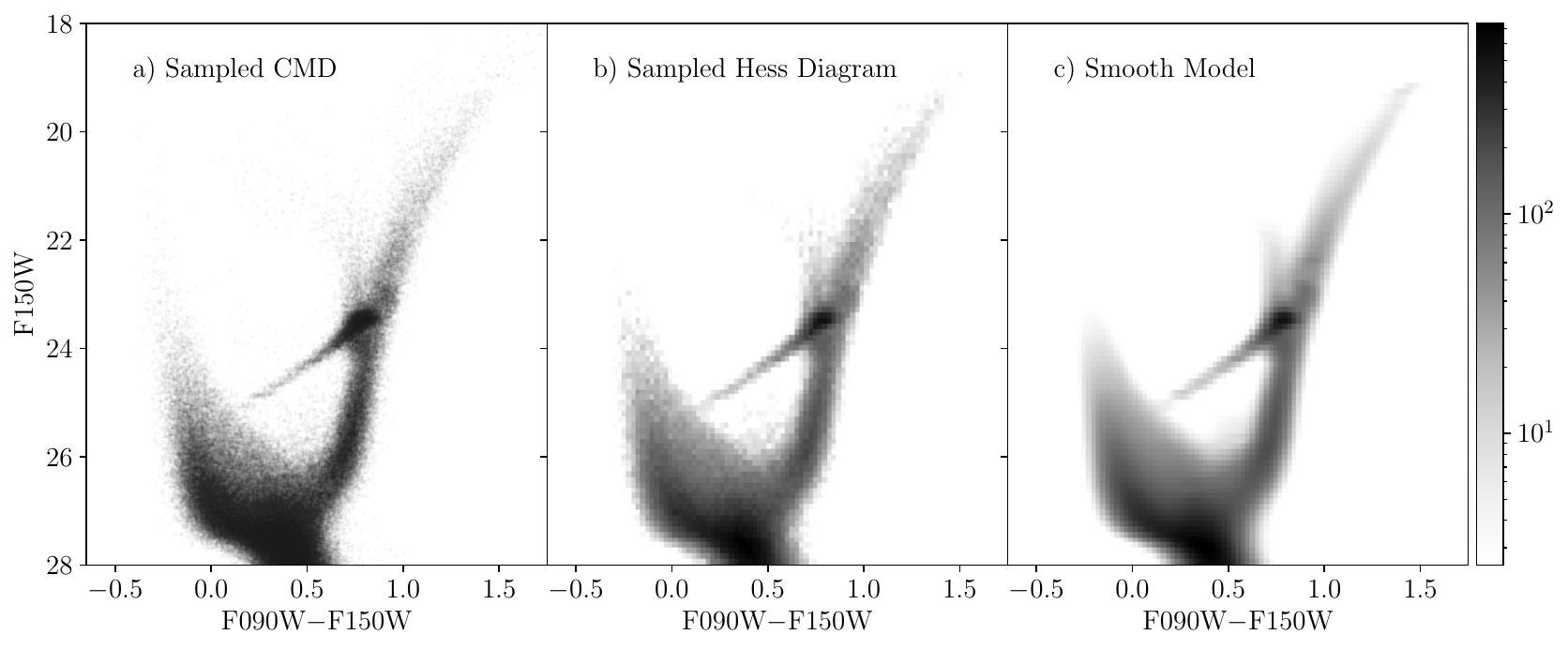}
  \caption{The CMD, Hess diagram, and optimal model for the synthetic stellar population constructed in \S \ref{sec:synthetic_data}. The properties of the synthetic population (e.g., $\text{M}_*=10^7$ M$_\odot$, $\mu=25$ mag) were chosen to roughly emulate WLM.}
  \label{figure:synthetic_cmd}
\end{figure*}

\begin{figure*}
  \centering
  \includegraphics[width=0.8\textwidth,page=1]{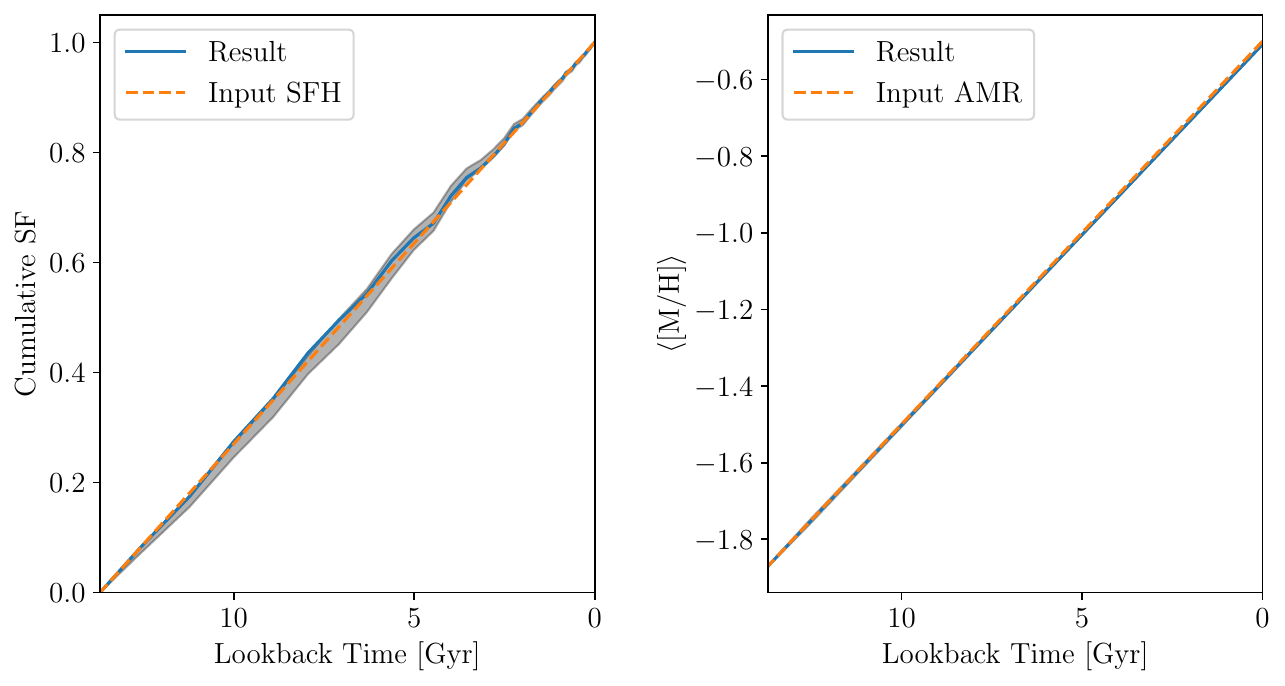}
  \caption{Results of applying our SFH fitting methodology to the synthetic stellar population shown in Figure \ref{figure:synthetic_cmd}. The best-fit cumulative SFH (left) and age-metallicity relation (right) are consistent with those used to construct the synthetic population. The $68\%$ credible intervals, considering only random uncertainties, are shaded in both panels, but this region is too small to see in the right panel.}
  \label{figure:synthetic_cumsfh}
\end{figure*}

\begin{figure*}
  \centering
  \includegraphics[width=0.4\textwidth,page=1]{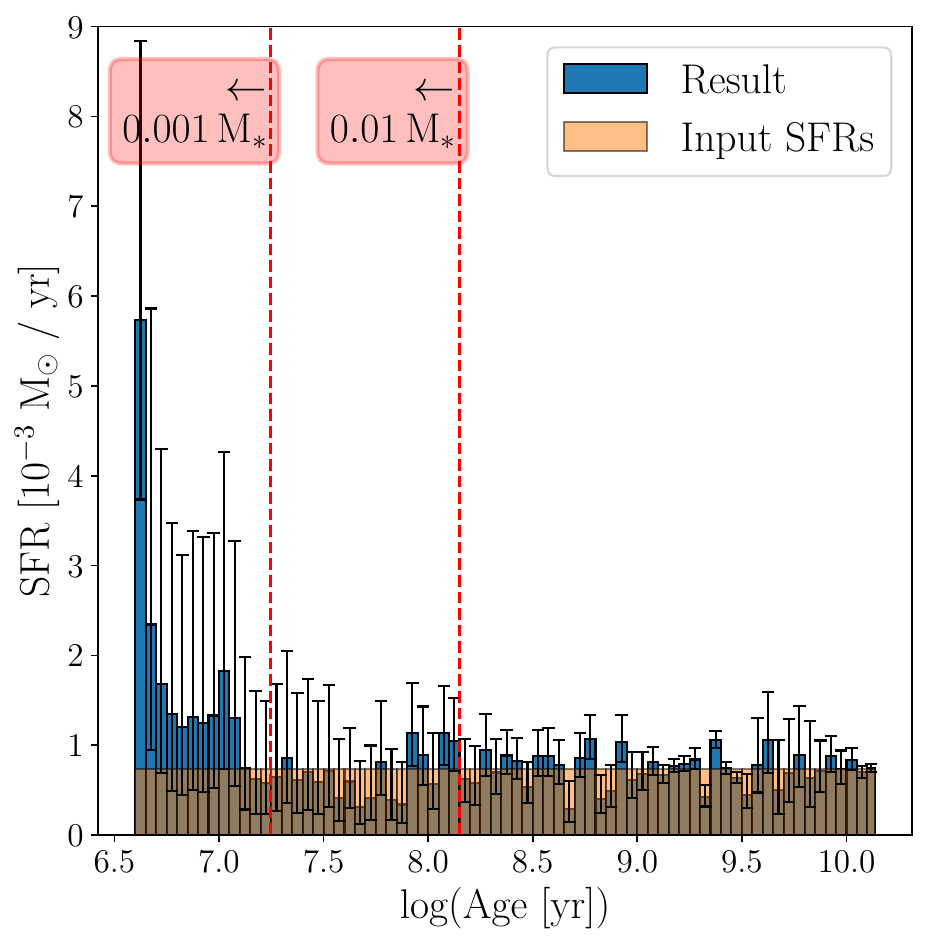}
  \hspace{1cm}
  \includegraphics[width=0.4\textwidth,page=1]{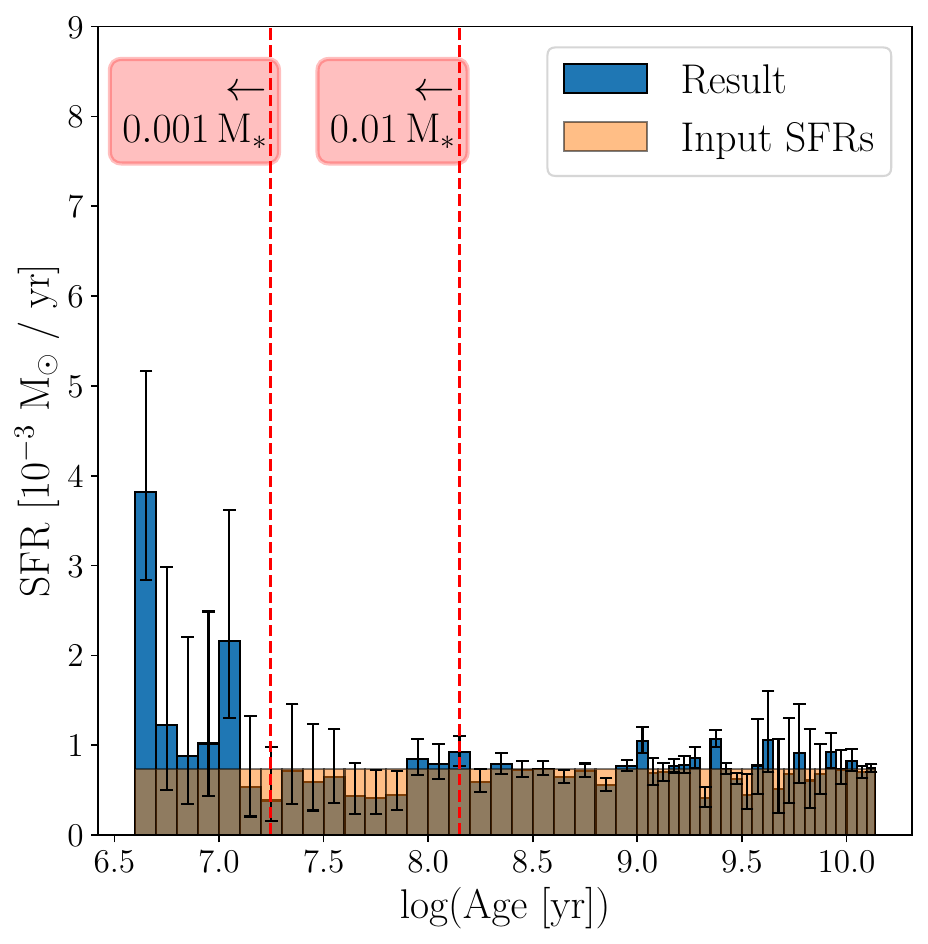}
  \caption{\emph{Left}: Best-fit star formation rates (SFRs) for the synthetic population shown in Figure \ref{figure:synthetic_cmd} compared to the constant intrinsic SFR of $\sim0.7\times10^{-3}$ M$_\odot$ yr$^{-1}$. Error bars show the $68\%$ credible intervals. There is very little stellar mass ($<0.01$ M$_*$) in the most recent time bins ($\log{ \left( \text{age} \right)} < 8$), which is why they have larger random uncertainties than older bins. We find good statistical agreement between the input SFRs and those we measure from the synthetic population. \emph{Right}: Same as left, but solved with a lower resolution grid for young stellar populations, resulting in reduced random uncertainties per SFR bin.}
  \label{figure:synthetic_sfrs}
\end{figure*}

\section{Application to Synthetic Data} \label{sec:synthetic_data}

Here we will use the CMD modelling techniques described in the previous section to measure the SFH of a synthetic stellar population constructed from theoretical isochrones. This is useful as an internal consistency test, as the myriad of systematics that appear in real data can be ignored and the performance of the underlying methodology can be assessed with perfect knowledge of the ground truth. The Jupyter notebook containing the analysis for this section is available to view in our source code repository.

Many different choices could be made with respect to the type of synthetic stellar population to test. Given that one of the main ways our approach differs from others is how we choose to model the metallicity evolution of the population, a synthetic population with a long history of star formation that results in a broad integrated MDF would be the best test of our approach. Such a population would be difficult to model without a way to constrain the set of possible metallicity evolutions, like our hierarchical model setup (\S \ref{subsec:hierarchical}). Comparatively, a UFD-like population with a very short period of star formation and relatively little metallicity evolution would be a less effective test of our metallicity distribution model. Additionally, we apply our methodology to HST/ACS data of the Horologium I UFD in \S \ref{sec:UFDs} and show that we are able to reproduce previous literature results.

Given this objective, we choose to construct a synthetic population reminiscent of WLM to perform our testing. As shown in \cite{McQuinn2024} and \S \ref{sec:wlm}, WLM has been actively star-forming over the majority of cosmic time and has a complex CMD morphology which is challenging to model. This choice will also provide a frame of reference for interpreting our results on the real data of WLM in \S \ref{sec:wlm}.

For simplicity, we model our synthetic population as having a stellar mass of $10^7$ M$_\odot$ (\citealt{McQuinn2024} measure $\log{\left(\text{M}_*\right)}=7.12$ within the JWST/NIRCAM field of view) and constant star formation from a lookback time of 13.7 Gyr to the present-day, resulting in a SFR of $\sim0.00073$ M$_\odot$ yr$^{-1}$. We impose an AMR of the form $\langle[\text{M}/\text{H}]\rangle(t)=\alpha \left( T_{max} - t \right) + \beta$ with $\alpha=0.1$, $\beta=-1.87$, and $T_{max}=13.7$ Gyr. We set the width of the Gaussian MDF at fixed time that appears in Equation \ref{eq:hmodel} to be $\sigma=0.1$ dex.

To construct the ``pure'' catalog of stars, we adopt a dense grid of PARSEC \citep{Bressan2012,Chen2014,Tang2014,Chen2015,Marigo2017,Chen2019a,Pastorelli2019,Pastorelli2020} isochrones with $\log{ \left( \text{age} \right) }$ ranging from 6.6--10.10 with a spacing of 0.05 dex and [M/H] ranging from $-2.8$ to 0.3 with a spacing of 0.1 dex, totalling 2,059 distinct isochrones. This is moderately denser than the grid used to measure the SFH of WLM by \cite{McQuinn2024}. Stars are sampled from these isochrones according to the methods in Appendix \ref{appendix:montecarlo} using the \cite{Kroupa2001} IMF.

This ``pure'' catalog is then degraded to approximate real data by emulating photometric error and incompleteness. We base our models for the photometric error and completeness functions on those derived from the JWST/NIRCAM data discussed in \S \ref{sec:wlm} and model our population with a distance modulus $\mu=25$ mag ($d=1$ Mpc), which is consistent with the measured distance of WLM \citep[$\mu=24.93\pm0.09$ mag,][]{Albers2019}. About 180,000 stars with F150W$<28$ mag are left in our final synthetic catalog. We show this synthetic stellar population in Figure \ref{figure:synthetic_cmd}. The result of this process is a synthetic stellar population that shares many CMD features with WLM (e.g., a prominent upper MS, HB, red clump, and blue loop) but lacks observational systematics like foreground/background contamination. 

To measure the SFH of this synthetic population, we first construct smooth SSP templates using the methods described in \S \ref{subsec:templates} for all 2,059 isochrones in the grid used to create the population. We use the same inputs to construct the templates as were used to construct the synthetic population (e.g., photometric error and completeness functions, IMF, etc.). We then fit the 71 SFRs and three metallicity-related parameters ($\alpha$, $\beta$, $\sigma$) simultaneously using the BFGS algorithm. As we find that none of the best-fit SFRs are zero, we choose to derive random uncertainties on the fit parameters using the inverse Hessian method described in \S \ref{subsec:fitting} rather than more expensive sampling methods for this example.

As the SFRs in adjacent time bins can be highly correlated \citep{Dolphin2002, McQuinn2010a}, it is common to examine the cumulative SFH, defined as the fraction of stellar mass formed at times earlier than $t$. Such a cumulative statistic has the effect of canceling out the correlations between adjacent time bins that can be significant when looking directly at SFRs. The left panel of Figure \ref{figure:synthetic_cumsfh} shows our best-fit cumulative SFH for this synthetic population with the $68\%$ credible interval shaded. Our result is statistically consistent with the intrinsic SFH of the population. The right panel shows our best-fit AMR for the synthetic population, which is also in very good agreement with the intrinsic AMR. The $68\%$ credible interval is also shaded in the right panel, but it is too narrow to see.

The most direct probe of the fit quality, the derived SFRs, are shown in the left panel of Figure \ref{figure:synthetic_sfrs}. Statistically, the fit SFRs are in good agreement with the intrinsic constant SFR of $0.7\times10^{-3}$ M$_\odot$ yr$^{-1}$, but some additional interpretation is warranted. In particular, our choice to fit the synthetic population with a dense isochrone grid with a spacing defined in $\log{ \left( \text{age} \right)}$ results in some fit characteristics which benefit from additional explanation.

As we have chosen to use a constant input SFR and an isochrone grid with a logarithmic time spacing, the widths of the SFR bins in yr increase exponentially with the age of the population. That is, a bin that starts at $\log{ \left( \text{age} \right)}=6.6$ and ends at $\log{ \left( \text{age} \right)}=6.65$ has a width in years of only 0.48 Myr, while a bin that starts at $\log{ \left( \text{age} \right)}=10$ and ends at $\log{ \left( \text{age} \right)}=10.05$ has a width in years of 1.2 Gyr. Under a constant SFR, this means that the total stellar mass formed in each bin increases exponentially with increasing $\log{ \left( \text{age} \right)}$. For example, only $0.1\%$ of the stellar mass of the population was formed more recently than $\log{ \left( \text{age} \right)}<7.25$ and $1\%$ of the stellar mass was formed more recently than $\log{ \left( \text{age} \right)}<8.14$.
This is the main reason why the uncertainties on the SFRs at recent times ($\log{ \left( \text{age} \right)}<8$) have greater random uncertainties than the SFRs at earlier times.

Generally, reducing the number of unique time bins used to fit the SFRs results in lower random uncertainties per bin, as the number of free parameters in the fit is reduced and the amount of stellar mass allocated to each remaining bin is increased, potentially increasing the signal-to-noise ratio per bin significantly \citep{Dolphin2002}. This effect can be seen by comparing our fiducial SFR measurements in the left panel of Figure \ref{figure:synthetic_sfrs} to the right panel, which shows our SFR measurements for the same synthetic population using a wider $\Delta \log{ \left( \text{age} \right)}=0.1$ dex bin spacing for young stellar populations. The coarser grid spacing in the right panel is chosen to match that used to measure the SFH of WLM in \cite{McQuinn2024} and provides a good mix of precision and time resolution. It should be noted that use of a solution grid with age spacing that is \emph{too} coarse results in discontinuities between the SSP models in the Hess diagram, leading to poor fitting residuals and potential systematic fitting errors. An often useful middle ground can be found by fitting with a reasonably high time resolution and statistically combining adjacent SFR bins after sampling if lower random uncertainties per bin are desired.

We also wish to note that at early times ($\log{ \left( \text{age} \right)}>9$) the SFRs exhibit significant correlation between adjacent time bins with Pearson correlation coefficients as significant as $-0.6$. This is a result of the fact that the CMD of two old populations with only a small difference in age can be morphologically very similar and could, therefore, fit the observed population similarly well. These correlations cancel out in the cumulative SFH, but they are significant when looking directly at SFRs. This produces a pattern where bins with best-fit SFRs that are higher than the intrinsic value are often neighbored by bins with best-fit SFRs that are lower than the intrinsic value. These covariances are present in the samples we draw to derive our uncertainties and can be visualized with corner plots or measured quantitatively by calculating the correlation matrix of the samples, but they are not obvious when looking at the SFRs in one dimension as in Figure \ref{figure:synthetic_sfrs}.

In summary, we are able to accurately recover the input SFH for our synthetic population. This is most obvious in the best-fit cumulative SFH and AMR shown in Figure \ref{figure:synthetic_cumsfh}, as these are less affected by specifics of the analysis (e.g., the time resolution of the adopted isochrone grid) than the SFRs shown in Figure \ref{figure:synthetic_sfrs}. Even so, the best-fit SFRs show good statistical agreement with the input value of $\sim0.7\times10^{-3}$ M$_\odot$ yr$^{-1}$, giving us confidence to apply our methodology on real data. We include synthetic data experiments in Appendix \ref{appendix:sfh_examples} that illustrate the performance of the SFH measurements on more distant galaxies without resolved MSTOs. We additionally examine the impact of observational systematics (particularly uncertainties on distances, foreground extinctions, and stellar binary fractions) on our ability to recover cumulative SFHs and AMRs in Appendix \ref{appendix:obs_systematics}.

\begin{figure*}
  \centering
  \includegraphics[width=0.75\textwidth,page=1]{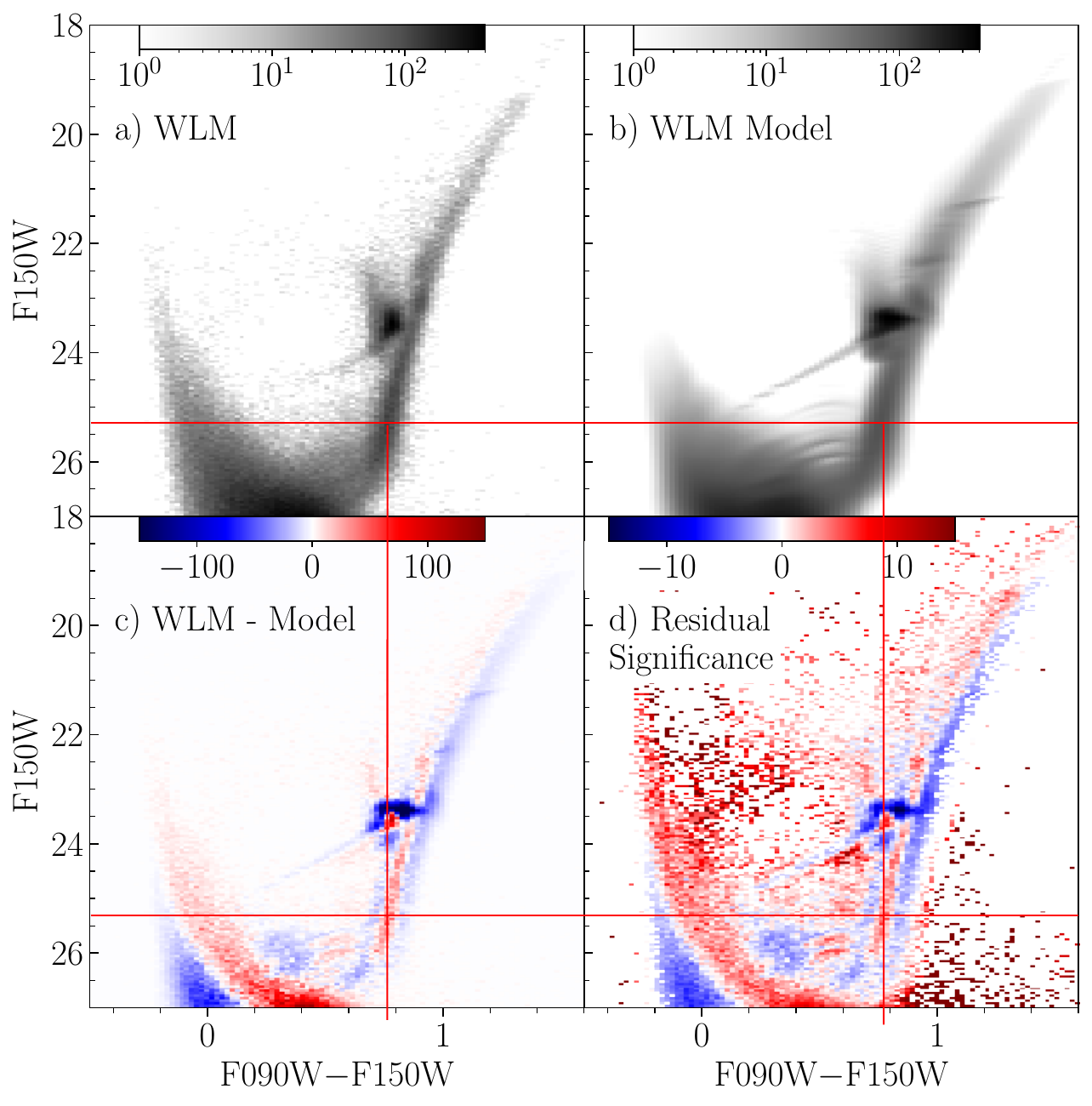}
  \caption{Comparison of WLM data to our best-fit model, formatted for easy comparison to figure 6\emph{a}) of \cite{McQuinn2024} who measured the resolved SFH of WLM with \textsc{match}. \emph{a}) Hess diagram of WLM, based on the catalogs of \cite{Weisz2024} for the full JWST/NIRCAM field of view. \emph{b}) Optimal model Hess diagram. \emph{c}) Residual between the observed and model Hess diagram in raw star counts. \emph{d}) Residual between the observed and model Hess diagram in units of standard deviations (i.e., the residual significance).}
  \label{figure:wlm_hess}
\end{figure*}

\begin{figure*}
  \centering
  \includegraphics[width=0.75\textwidth,page=1]{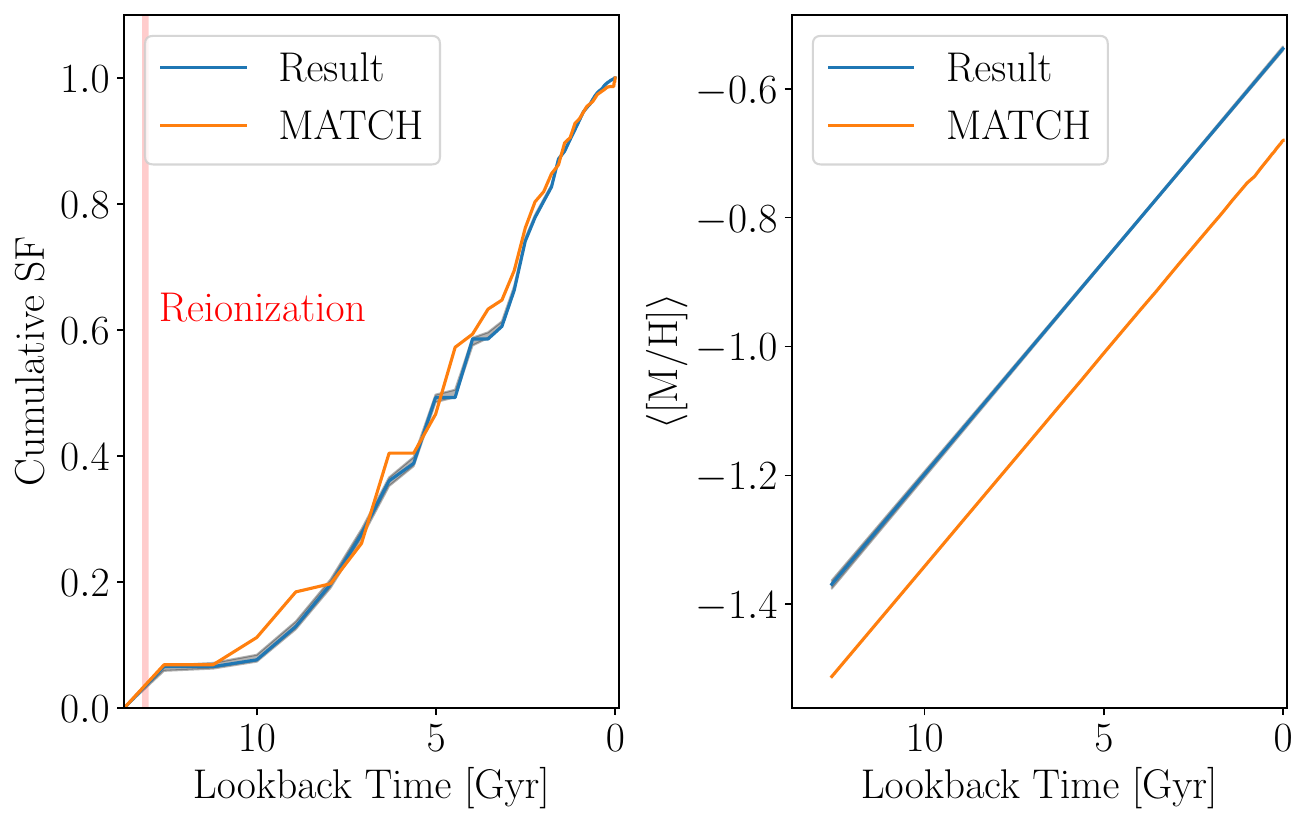}
  \caption{\emph{Left}: Cumulative SFHs of WLM measured from the JWST/NIRCAM photometric catalogs of \cite{Weisz2024} with our methodology (blue) and \textsc{match} \citep[orange,][]{Cohen2025}. The $68\%$ credible interval for our result, considering only random uncertainty, is shaded grey. While there are minor differences, the two results are generally in good agreement. The shaded red region shows the $68\%$ confidence interval for the midpoint of reionization \citep{Planck2020}. \emph{Right:} Solutions for the age-metallicity relation found with our methodology (blue) and \textsc{match} \citep[orange,][]{Cohen2025}. The slopes are in excellent agreement, with only a minor $\sim0.1$ dex constant offset. This is likely due to differences in how we model the fixed-age metallicity distribution.}
  \label{figure:wlm_cumsfh}
\end{figure*}

\section{WLM} \label{sec:wlm}
WLM is a special galaxy in the Local Group as it is a dwarf irregular \citep[$M_v\approx-14.2$, $\text{M}_*\approx4.3\times10^7$ M$_\odot$,][]{McConnachie2012} that is isolated and gas-rich. With a distance modulus of $\mu=24.93\pm0.09$ mag \citep[$d=968_{-40}^{+41}$ kpc,][]{Albers2019}, it is near enough that deep HST/ACS and JWST/NIRCAM imaging have enabled measurements of its resolved SFH with excellent time resolution \citep{Albers2019,McQuinn2024}. These results have shown that, as expected for a gas-rich, isolated dwarf galaxy, WLM has been forming stars throughout most of cosmic history, though its average SFR has increased in the last $\sim7$ Gyr. This is partly due to WLM having very little star formation activity in the $\sim3$ Gyr following reionization. The work on WLM by \cite{McQuinn2024} was also the first to demonstrate such high-precision resolved SFH measurements in the infrared with JWST/NIRCAM, which is likely to supercede HST/ACS as the preferred data source for these types of resolved SFH studies in the future.

With a rich legacy of work in the literature and publicly available photometric catalogs and ASTs for both HST/ACS \citep{Albers2019} and JWST/NIRCAM \citep{Weisz2024}, WLM is an excellent test case for our methodology. Being massive compared to dwarf satellites of the MW, WLM displays a complex but well-sampled CMD for which the random errors in the resolved SFHs are extremely low. A comparative study of WLM should therefore reveal any systematic differences between \textsc{match} and our methodology, as long as ``input'' systematics (e.g., assumed distance, interstellar reddening, and stellar models) are the same.

Since \cite{McQuinn2024} used pre-release photometric catalogs that are not publicly available and focused their analysis on the region of WLM where the HST/ACS and JWST/NIRCAM imaging overlap, we do not compare directly to their published results. Rather, we compare against new \textsc{match} results that use the public photometric catalogs and ASTs released as part of the JWST Resolved Stellar Populations Early Release Science Program \citep{Weisz2024} for the full JWST/NIRCAM field of view \citep{Cohen2025}. The modelling assumptions and \textsc{match} parameters used to make these measurements are the same as were used in \cite{McQuinn2024}, with the only difference being the input photometric catalogs and ASTs. These results are qualitatively consistent with those published in \cite{McQuinn2024}, as differences between the pre-release catalogs used by \cite{McQuinn2024} and the public catalogs are minimal. We use the same public photometric catalogs \citep{Weisz2024} and both analyses apply the star-galaxy separation criteria from \cite{Warfield2023}. As \cite{McQuinn2024} demonstrated that the resolved SFHs derived from the HST/ACS data and the JWST/NIRCAM data are consistent, we consider only the JWST/NIRCAM data here. We apply our methodology to HST/ACS data of the Horologium I UFD in \S \ref{sec:UFDs} so that we have an example of our methodology applied to both observatories.

We assume the following parameters in our fit for WLM, mirroring the choices made to derive the \textsc{match} result:
\begin{enumerate}
  \item $A_V = 0.1$ mag from the dust maps of \cite{Schlegel1998} with the updated scaling from \cite{Schlafly2011}.
  \item Distance modulus $\mu=24.93\pm0.09$ mag \citep[$d=968_{-40}^{+41}$ kpc,][]{Albers2019}.
  \item PARSEC v1.2S isochrones \citep{Bressan2012,Chen2014,Tang2014,Chen2015,Marigo2017,Chen2019a,Pastorelli2019,Pastorelli2020} with scaled-solar abundance patterns.
  \item Binary fraction $=35\%$, added with the method described in \S \ref{subsec:binaries}.
  \item \cite{Kroupa2001} IMF.
\end{enumerate}
To illustrate the performance of the hierarchical AMR models developed in \S \ref{subsec:hierarchical}, we adopt the linear [M/H]($t$) model and fit its slope and intercept simultaneously with the SFH. We fix the metallicity dispersion at fixed time (the $\sigma$ that appears in Equation \ref{eq:hmodel}) to 0.2 dex.

The left panel of Figure \ref{figure:wlm_hess} shows the JWST/NIRCAM Hess diagram of WLM, our best-fit smooth model, and the model residuals in raw counts and in units of standard deviations (i.e., the residual significance). These plots are designed to facilitate easy comparison to figure 6\emph{a}) of \cite{McQuinn2024} -- their plot is made using only data in the region of overlap between the HST/ACS and JWST/NIRCAM data, so there are fewer stars overall than in our plot that uses the full JWST/NIRCAM field of view, but we have made the bins in our Hess diagram smaller so that we achieve similar signal-to-noise per bin. This enables us to show the residual significance values in panel \ref{figure:wlm_hess}\emph{d}) on the same scale as figure 6\emph{a}) of \cite{McQuinn2024}. The pattern and magnitude of residuals in our fit are extremely similar to theirs, with the following features in common:
\begin{enumerate}
  \item An excess of stars in the model along the red side of the RGB.
  \item Large residuals around the red clump.
  \item An excess of observed stars on the red side of the upper MS (e.g., $x=0, y=24$).
  \item An excess of stars in the model on the blue side of the MS at faint magnitudes (e.g., $x=0, y=26.5$).
\end{enumerate}
This consistency is reflected in the measured cumulative SFH shown in the left panel of Figure \ref{figure:wlm_cumsfh}. While there are minor differences at intermediate ages (e.g., 3--5 Gyr), our measurement is broadly consistent with that from \textsc{match} \citep{Cohen2025}. In particular, agreement is excellent within the last 3 Gyr, and we also replicate the quiescent period found by \textsc{match} following reionization \citep[$z_{re}=7.67\pm0.73$, $t_{re}=13.11\pm0.09$ Gyr,][]{Planck2020}. A useful comparison is figure 9 of \cite{McQuinn2024} which shows cumulative SFHs for WLM fit with different data (JWST/NIRCAM, HST/ACS, and a joint fit) and different stellar models -- the variance between these results is significantly greater than the discrepancies we see between our result and the \textsc{match} result in Figure \ref{figure:wlm_cumsfh}.

The right panel of Figure \ref{figure:wlm_cumsfh} compares the \textsc{match} AMR to our best-fit linear AMR of $\langle[\text{M}/\text{H}]\rangle(t)=0.07 \left( 13.7 - t \right) - 1.44$ where $t$ is the lookback time in units of Gyr. This slope is consistent with the \textsc{match} result while the intercept is $\sim0.1$ dex larger, likely due differences in how we model the fixed-age metallicity distribution. In any case, 0.1 dex is quite minor as metallicity inference from broadband photometry (particularly with only one color) is typically imprecise.

\begin{figure}
  \includegraphics[width=0.45\textwidth,page=1]{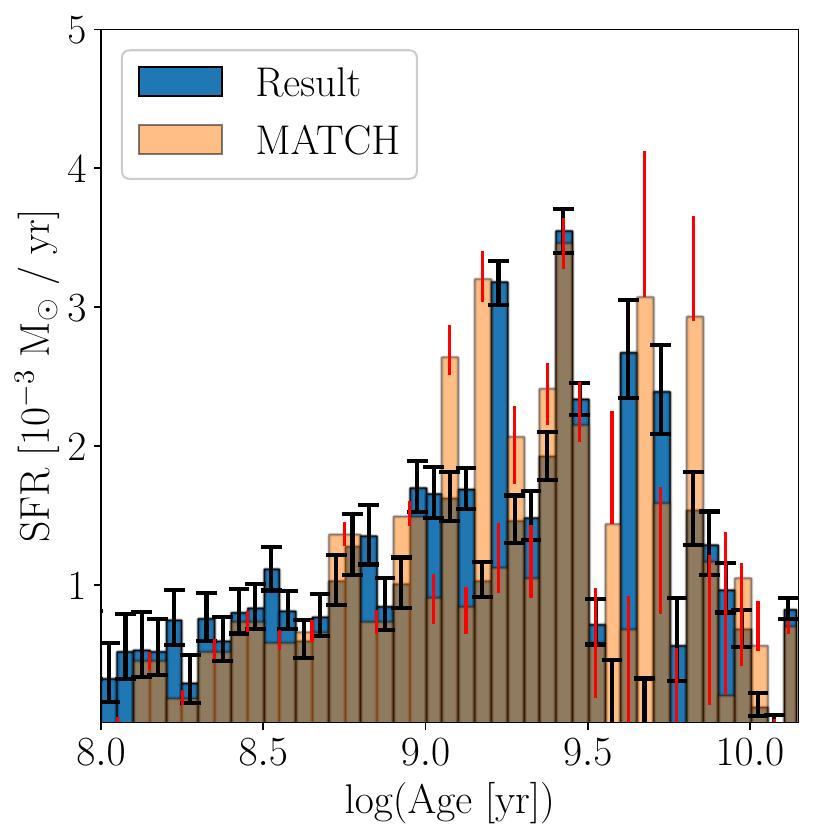}
  \caption{The global SFH of WLM measured from the JWST/NIRCam data with our method (blue bars, black error bars) compared to the measurement on the same data with \textsc{match} \citep[orange bars, red error bars;][]{Cohen2025}. Error bars shown are $68\%$ credible intervals considering only random (statistical) uncertainty. Agreement is excellent from the present-day to 1 Gyr ago ($x=9$) and good from 1 Gyr ago to about 4 Gyr ago ($x=9.6$). We see large covariances for time bins from 4 -- 6 Gyr ago ($x=$9.6--9.8) that result in worse agreement with the \textsc{match} result. We attribute these differences to the significant red clump features present in stellar populations of these ages under our fiducial AMR -- the red clump is known to be difficult to model to the level of precision achieved by our data.}
  \label{figure:wlm_sfr}
\end{figure}

While we recover a cumulative SFH for WLM that matches well with the qualitative features of the \textsc{match} result, it is instructive to consider what quantiative differences may exist more closely. For this purpose, we present the raw SFH of WLM in Figure \ref{figure:wlm_sfr} to supplement the cumulative SFH shown in Figure \ref{figure:wlm_cumsfh}. The measurement from \textsc{match} is also shown for comparison. We note that the \textsc{match} solution for times more recent than 1 Gyr has a time resolution of $\Delta \log{ \left( \text{age} \right)}=0.1$ dex and changes to $\Delta \log{ \left( \text{age} \right)}=0.05$ dex for earlier times, while our solution uses $\Delta \log{ \left( \text{age} \right)}=0.05$ dex throughout. The \textsc{match} uncertainties (per age bin) are therefore smaller than ours for times more recent than 1 Gyr. The effect of reducing the time resolution on the SFR measurement uncertaintes is shown in Figure \ref{figure:synthetic_sfrs}.

For times more recent than 1 Gyr $\left(\log{ \left( \text{age} \right)} < 9\right)$ we see very good agreement between our measurement and that from \textsc{match} which we attribute to the well-sampled upper MS allowing for a robust measurement. Agreement is good from 1 -- 4 Gyr ago $\left(9 \leq \log{ \left( \text{age} \right)} \leq 9.6\right)$, with some SFR transposition due to temporal covariance. As previously mentioned, SFRs in adjacent time bins can be highly correlated \citep{Dolphin2002, McQuinn2010a}, an effect which is mitigated in a cumulative statistic like the cumulative SFH. In the raw SFRs we are now considering, these covariances do not cancel out, but cannot be easily visualized either -- the error bars are typically standard deviations and so do not convey these covariances. The SFRs in the range of 1 -- 4 Gyr ago exhibit the effects of this covariance, where under (over) estimates of the SFR relative to \textsc{match} are generally offset by over (under) estimates in an adjacent bin.

The only time period where we see divergence from \textsc{match} not explained by inter-bin covariance is in the period from 4 -- 6 Gyr ago $\left(9.6 \leq \log{ \left( \text{age} \right)} \leq 9.8\right)$. Additionally, both solutions show a large degree of time variability in the SFR during this time period. We believe these are related -- populations with ages 4 -- 6 Gyr under our fiducial AMR model have significant red clump features, and while the red clump of WLM is very well-populated, it is difficult to model the red clump to the precision of space-based observatories like HST/ACS \citep[see, e.g., Figure 4 of][]{Albers2019} and JWST/NIRCam \citep[see, e.g., Figure 6a of][and our Figure \ref{figure:wlm_hess}]{McQuinn2024}. As such, we hypothesize that the discrepancies in the SFRs 4 -- 6 Gyr ago are driven by difficulties modelling the red clump features of populations with these ages. In such cases where particular CMD features present modelling challenges, a CMD discretization technique that can separate the CMD by feature \citep[such as that used by IAC-pop;][]{Aparicio2009} would be useful. One could imagine a scheme by which the MS and subgiant branch may be given increased statistical weight in the fit relative to evolved star features (e.g., the red clump and horizontal branch) could mitigate some of these challenges. We leave exploration of this to future work.

Again, the cumulative SFH we measure for WLM is in good agreement with the results from \textsc{match} \citep{McQuinn2024,Cohen2025}, illustrating all the same qualitative features. Quantitatively, the raw SFRs we measure are also in good agreement with \textsc{match} across most of cosmic time, with the exception of a small period of divergence 4 -- 6 Gyr ago which we attribute to challenges in modelling the red clump of these populations at the level of precision achieved by our JWST/NIRCam photometry. These results give us confidence in our methodology and form an important point of reference for comparing results obtained with our methodology to those measured with \textsc{match}, which has been used to make many key measurements in the field.


\begin{figure*}
  \centering
  \includegraphics[width=0.75\textwidth,page=1]{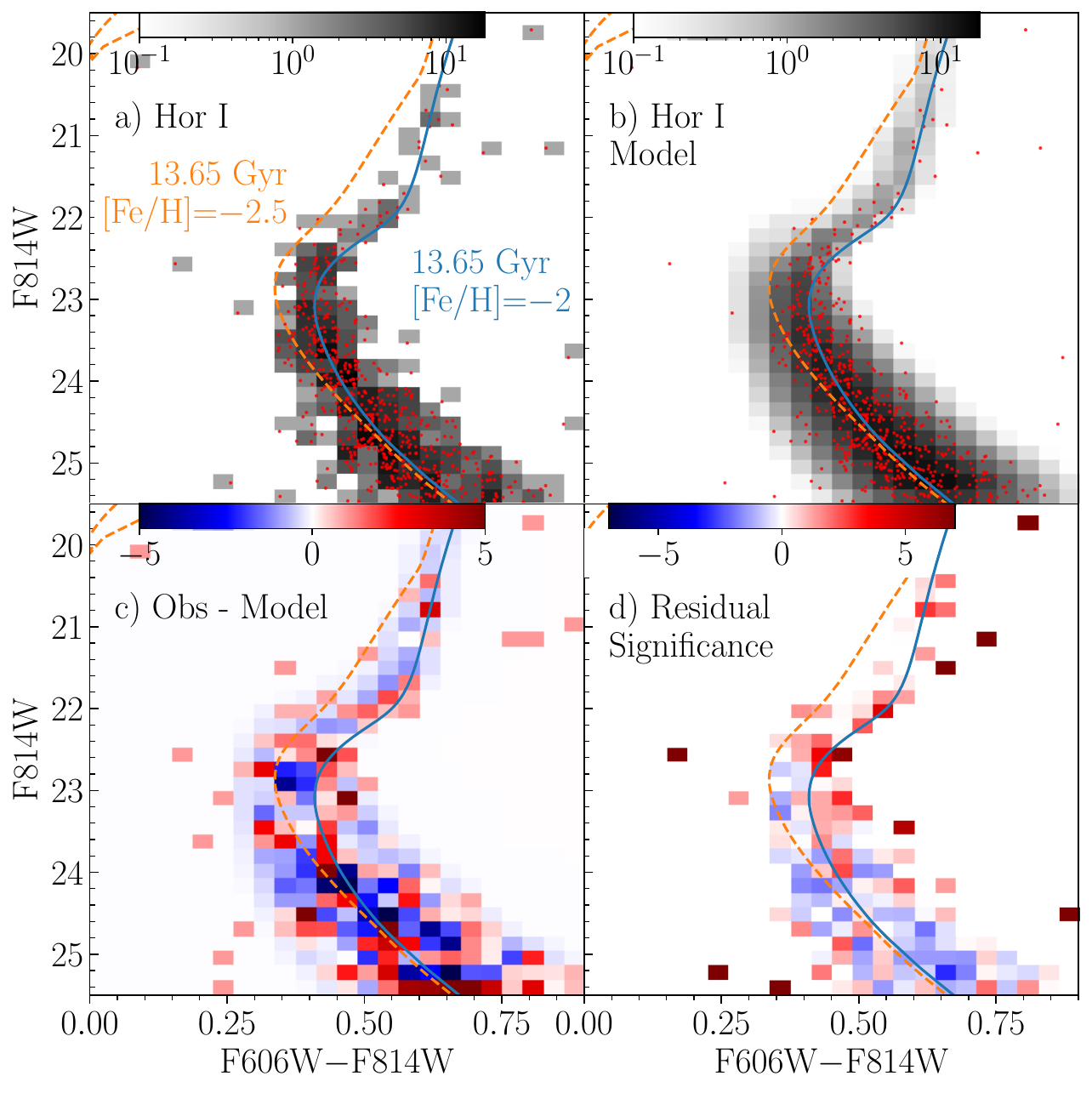}
  \caption{Analog of Figure \ref{figure:wlm_hess} for Hor I. Our best-fit model with $\text{[Fe/H]} = -2$ provides a good fit to the data, but observational constraints on the iron abundance suggest it should have $\text{[Fe/H]} \leq -2.6$. We find the MIST isochrones with such low iron abundances have MSTO colors that are too blue to provide adequate fits. We show 13.65 Gyr isochrones with both $\text{[Fe/H]} = -2$ (blue solid line) and $-2.5$ (orange dashed line) to illustrate this.}
  \label{figure:hor_i_hess}
\end{figure*}

\begin{figure}
  \includegraphics[width=0.45\textwidth,page=1]{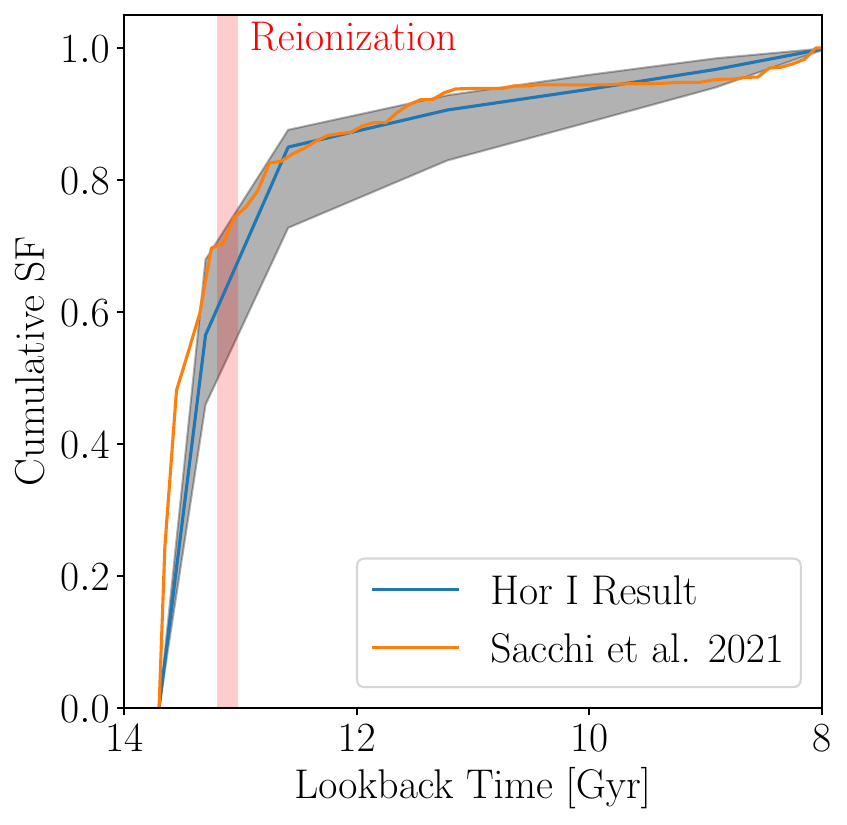}
  \caption{The cumulative SFHs of the Horologium I UFD measured from the HST/ACS aperture photometry of \cite{Richstein2024} with our methodology (blue line) and \textsc{sfera} \citep[orange line,][]{Sacchi2021}. The $68\%$ credible interval for our result, considering only random uncertainty, is shaded grey. The shaded red region shows the $68\%$ confidence interval for the midpoint of reionization \citep{Planck2020}. As Horologium I is ultra-faint $M_v\approx-3.7$, its CMD is sparsely populated, resulting in larger random uncertainties than we saw for WLM in Figure \ref{figure:wlm_cumsfh}. Agreement is very good between our result and that of \cite{Sacchi2021}.}
  \label{figure:hor_i_cumsfh}
\end{figure}

\section{Horologium I} \label{sec:UFDs}
While the previous section illustrated the performance of our methods on a complex stellar population with a well-sampled CMD, it is also important to examine how well our methods perform on the sparsely-populated CMDs typical of UFDs. We will do so by measuring the resolved SFH of the Horologium I UFD and comparing to the measurement of \cite{Sacchi2021}, who used the \textsc{sfera} code \citep{Cignoni2015,Cignoni2019} with the MIST stellar models \citep{Choi2016,Dotter2016}. This is prefaced by a discussion of the challenges presented when measuring the resolved SFHs of UFDs to provide context for our modelling choices and results.

\subsection{Technical Considerations}
All previously studied UFDs host ancient stellar populations of ages $\geq10$ Gyr \citep{Brown2014,Weisz2014a,Weisz2014b,Sacchi2021,Savino2023}, which, in combination with their intrinsically low luminosities ($M_V>-7$ mag), makes them uniquely challenging to model. While their stellar populations are ancient, UFDs require a spread in stellar ages to explain their CMDs as measured with high-precision space-based observatories; i.e., they are not mono-age populations. Additionally, they have significant intrinsic metallicity spreads of 0.2--0.7 dex \citep[e.g.,][]{Kirby2011,Fu2023}. Given these factors, the CMDs of these objects are best analyzed with a SFH fitting code rather than simpler isochrone fitting techniques often applied to globular clusters \citep[e.g.,][]{VandenBerg2013,VandenBerg2022a}. As previously mentioned, the SFRs of old stellar populations can often have strong covariances due to how few age-sensitive CMD features there are in these old stellar populations. The primary CMD feature that provides age sensitivity for old, metal-poor populations is the MSTO -- the ability to fit the resolved SFH of a UFD is therefore highly dependent on the presence of a well-resolved and reasonably-sampled MSTO.

Even when the observational data are sufficient to attempt a measurement, the process is further complicated by systematics -- given the high precision of the space-based photometry used for these measurements and the relative simplicity of the stellar populations, the width of the MSTO feature in the CMDs of UFDs can often be $\leq0.1$ mag \citep[see, e.g., figure 7 of][]{Brown2014}. As such, uncertainties in systematics like the interstellar reddening, distance modulus, and population metallicity can affect the measured age of the population at the level of several hundred Myr -- each of these systematic uncertainties contribute (at minimum) a few hundredths of a magnitude to the overall error budget, resulting in a systematic error in the MSTO magnitude that can be of the same order of magnitude as the width of the MSTO itself. And these are just the systematics on the observational side -- at these levels of precision, systematic uncertainties in the stellar models (e.g., the helium abundance and heavy-element mixture; see \citealt{VandenBerg2013}; \citealt{VandenBerg2022a}) are comparable to, if not greater than, the observational systematics. These systematics predominantly affect \emph{absolute} age measurements like $\tau_{50}$ (the lookback time prior to which $50\%$ of the present-day stellar mass was formed), as these systematics typically shift population-level quantities like the MSTO color. \emph{Relative} age measurements, like the duration of star formation, are more robust to these systematics and can be quantified through statistics like $\tau_{90} - \tau_{50}$.

It is therefore important when comparing measurements of the resolved SFHs of UFDs to consider differences in the assumptions made with respect to these systematics. When measuring the resolved SFH of WLM, we wanted to test our new constrained AMR model and so our metallicity model was different than that used in \textsc{match} by \cite{McQuinn2024}, whose result we compared against. Due to the complexity of the SFH of WLM, its CMD features are much broader than those of UFDs and these systematics do not affect the fit significantly. When looking at UFDs, these systematics will be much more significant. Therefore, in this section we will attempt to mirror the systematic choices made in \cite{Sacchi2021} to minimize differences that could arise due to the authors' choice of isochrone library, reddening map, metallicity, etc. Our goal here is to show that given the same observational data and systematic modelling assumptions, our method can produce results consistent with established literature methods for UFDs.

\subsection{Fitting the SFH of Horologium I}
For this experiment we were interested in comparing against the SFH measurements for UFDs presented in \cite{Sacchi2021} as they employ the original version of \textsc{sfera} \citep{Cignoni2015,Cignoni2019} to make their measurements, which used Monte Carlo sampling to generate templates, giving us another point of reference in addition to our comparison with \textsc{match} for WLM. Additionally, \cite{Sacchi2021} used the MIST stellar models \citep{Choi2016,Dotter2016}, while we used the PARSEC models when analyzing WLM, giving us an opportunity to examine how our method performs with a different stellar library.

In \cite{Sacchi2021}, the authors measure the resolved SFHs of six UFDs with $-5.2 < M_V < 1.8$ and distances $31 < d < 150$ kpc using preliminary aperture photometry catalogs derived from the imaging data presented in \cite{Richstein2024}. We use these same catalogs in our analysis for consistency. For our experiment, we choose to study the Horologium I (Hor I) UFD \citep{Bechtol2015,Koposov2015} as it is intermediate in both luminosity ($M_V\approx-3.4$) and distance ($d \approx 83$ kpc) relative to the range of the overall sample. Both spectroscopic \citep{Koposov2015a,Nagasawa2018} and CaGK photometric \citep{Fu2023} iron abundance measurements show that Hor I has $-2.8 < \text{[Fe/H]} < -2.6$, which is lower than average for the sample of UFDs in \cite{Sacchi2021}. High-resolution spectroscopy of three stars in Hor I suggests that it may not be alpha-enhanced like many other UFDs ($[\alpha/\text{H}] \approx0$, \citealt{Nagasawa2018}). Hor I is also interesting as its dynamics suggest that it was a satellite of the LMC prior to its recent accretion into the MW's halo \citep{Kallivayalil2018} such that Hor I was likely in an environment less dense than the MW during reionization.

We assume the following parameters in our fit for Hor I, following \cite{Sacchi2021}:
\begin{enumerate}
  \item $A_V = 0.04$ mag from the dust maps of \cite{Schlegel1998} with the updated scaling from \cite{Schlafly2011}. 
  \item Distance modulus $\mu=19.6$ mag ($d\approx83$ kpc). This is the average of $19.7\pm0.2$ mag measured by \cite{Bechtol2015} and $19.5\pm0.2$ from \cite{Koposov2015}.
  \item MIST v1.2 isochrones \citep{Choi2016,Dotter2016} with scaled-solar abundance patterns.
  \item Binary fraction $=30\%$, added with the method described in \S \ref{subsec:binaries}.
  \item \cite{Kroupa2001} IMF.
\end{enumerate}
As we have already discussed, resolved SFH solutions for UFDs can be very sensitive to the assumed population metallicity as it can significantly affect the color of the MSTO.
The solution by \cite{Sacchi2021} optimized for the best-fit mean iron abundance ($\langle[\text{Fe}/\text{H}]\rangle$) in the range $-4 \leq \text{[Fe/H]} \leq -1$ with 0.1 dex steps. As we could not determine what they found to be the best-fit [Fe/H] for Hor I, we decided to perform the same optimization. While in principle we could include priors on the iron abundance based on observations, we found this to perform poorly. As discussed in \cite{Dolphin2016a}, it is not uncommon for the best-fit metallicities implied by broadband photometry to be inconsistent with spectroscopic metallicities due to systematics in the stellar models or bolometric corrections used to model the photometry.
\cite{Dolphin2016a} addresses this by fitting offsets between the spectroscopic and photometric metallicities as a function of stellar color, magnitude, and age, but we do not have sufficient numbers of spectroscopic observations to make this a viable strategy. We therefore place a uniform prior on the iron abundance of Hor I and find a best-fit value of $\text{[Fe/H]}=-2$, while spectroscopy \citep{Koposov2015a,Nagasawa2018} and narrow-band photometry \citep{Fu2023} favor values $-2.8 < \text{[Fe/H]} < -2.6$.

We show the Hess diagram of Hor I, our best-fit Hess diagram model, and the model residuals in Figure \ref{figure:hor_i_hess}. For comparison we show a very old isochrone with age 13.65 Gyr at [Fe/H] values of both $-2.5$ and $-2$. The isochrone for our best-fit value of $\text{[Fe/H]}=-2$ tracks the MS and MSTO well, though the isochrone is $\sim0.02$ mag too blue along the RGB. In comparison, the MSTO of the $\text{[Fe/H]}=-2.5$ isochrone is too blue by nearly $0.1$ mag, and even if we shift the isochrone to the correct MSTO color, its morphology is a poor fit to the observations. Younger isochrones at these iron abundances will have MSTOs that are bluer and brighter, moving towards the upper left of the plot, such that the fit for $\text{[Fe/H]}=-2.5$ becomes even worse for younger ages. This seems strange in light of the fact that \cite{Fu2023} used MIST models to measure a photometric iron abundance of $\text{[Fe/H]}\approx-2.8$ for Hor I, which is consistent with the spectroscopic results, but we hypothesize the difference may be due to systematics in the adopted bolometric corrections. \cite{Fu2023} used narrow-band CaHK (F395N) imaging taken with HST/UVIS to derive their photometric iron abundances, while we are modelling broadband photometry at slightly longer wavelengths. We suggest the bolometric corrections used in the MIST isochrones may accurately model the narrow-band CaHK (F395N) at lower iron abundances, explaining the \cite{Fu2023} result, while being less robust in the broadband F606W and F814W filters, explaining our result. Note as well that the $\text{[Fe/H]}=-2$ MIST isochrone shown overlaid on the CMDs in figure 1 of \cite{Sacchi2021} appears to match the MSTO colors and morphologies of the other UFDs in the sample fairly well, suggesting this iron abundance is reasonable when using the MIST isochrones.

Our best-fit cumulative SFH and $68\%$ credible interval are shown in Figure \ref{figure:hor_i_cumsfh} with the best-fit result of \cite{Sacchi2021} for comparison. The credible interval for our result is derived using the HMC sampling technique discussed in \S \ref{subsec:sampling}. The random uncertainties quoted by \cite{Sacchi2021} are comparable to ours (see their figure 3) such that our results are statistically consistent across the majority of cosmological history. We also measure consistent values of $\tau_{50}$ and $\tau_{90}$, which are the lookback times prior to which $50\%$ and $90\%$ of the present-day stellar mass was formed, respectively. We find $\tau_{50}=13.35^{+0.06}_{-0.15}$ Gyr and $\tau_{90}=11.12^{+0.88}_{-1.85}$ Gyr while \cite{Sacchi2021} find $\tau_{50}=13.44\pm0.17$ and $\tau_{90}=11.53\pm1.13$ (their table 2). Overall, these results illustrate that our method can be robustly applied to measure resolved SFHs of UFDs that are consistent with other literature methods.


\section{Discussion} \label{sec:discussion}

We have described the methods and design of \textbf{StarFormationHistories.jl}, our new open-source Julia package for fitting resolved SFHs, and demonstrated its application to JWST/NIRCAM of the dIrr WLM and HST/ACS data of the UFD Horologium I. This work has shown that our methodology can robustly measure resolved SFHs across a wide range of galaxy properties (e.g., stellar mass and SFH complexity). The modular design of the package supports arbitrary IMFs and allows other population parameters to be fit as hyperparameters (e.g., distance, binary fraction, IMF slope; Geha et al., in preparation). We leverage the strengths of the Julia programming language to achieve excellent runtime performance that has allowed us to run all the analysis for this work on personal computers. 

The example applications we have performed in this work demonstrate that the package is already capable of reproducing the types of resolved SFH analyses often done in the literature. However, there are several aspects of the package which we hope to improve or extend in the future to enable new insights. For example, our linear and logarithmic AMR models cover two common AMR morphologies but are not as flexible as a non-parametric method (as implemented in \textsc{match}) or as physically-motivated as a coupled chemical enrichment model \citep[as supported in IAC-STAR,][]{Aparicio2004}. We plan to extend the available AMR models in the future to provide greater flexibility. 

Another opportunity for development lies with closer integration with stellar models. We designed our package to be used with user-provided isochrones for extensibility and simplicity. However, access to the raw stellar tracks with their much higher time resolution would open up additional options in the template creation process. We could, for example, interpolate across tracks to introduce age spread into the templates themselves as is done in \textsc{match} \citep{Dolphin2002}. Direct access to the stellar tracks would also enable more options for estimating systematic uncertainties. Using the raw stellar tracks is non-trivial as libraries of stellar tracks are formatted differently, necessitating custom interfaces for each library, but the benefits are such that we consider supporting tracks a high priority.

We have so far neglected discussing systematic uncertainties because we have tried to make the same assumptions in our analyses as the literature work that we have been comparing to. However, for future, original analyses we will need to address systematics. For resolved SFHs, the dominant systematic uncertainties come from the stellar evolutionary tracks and bolometric corrections used to model the observations \citep{Gallart2005,Dolphin2012,Dolphin2016a}, which is why special attention should be paid to the stellar libraries used for these analyses. 
It is common to constrain these systematic uncertainties by measuring the resolved SFH from the same observational data with multiple stellar libraries \citep[e.g.,][]{Skillman2017,Sacchi2021}, the idea being that systematic differences in the underlying stellar tracks and BCs will propagate to systematic differences in the resolved SFH measurement. It is easy to use isochrones from different stellar libraries with our code, making this form of systematic uncertainty estimation simple. 

While this approach is certainly useful, the full range of the systematics cannot be sampled by considering a handful of different stellar libraries, as discussed in section 3 of \cite{Dolphin2016a}. In this work, the author develops an alternative methodology, implemented in \textsc{match}, that derives systematic uncertainties in the resolved SFH measurement by applying offsets to the effective temperatures and bolometric magnitudes of the fiducial stellar tracks to more fully sample the range of systematic uncertainties in the tracks. 
While this method can provide a more complete accounting of systematics than remeasuring the same data with different stellar libraries, determining the appropriate ranges within which to vary the effective temperature and bolometric magnitude can be challenging, as the offsets are meant to approximate the systematics in the stellar models which cannot be determined from first principles. We will investigate such methods for systematic uncertainty estimation after we add support for using libraries of stellar tracks and bolometric corrections.

The most precise constraints on galactic SFHs come from modelling high-precision CMDs, making the methodologies underlying these measurements critically important to the observational study of galaxy evolution. As we notice areas for improvement while applying StarFormationHistories.jl to more stellar populations in the local Universe and beyond, we will continually seek to improve and extend our methods to enable new and innovative analyses. By releasing our package as open-source software we make these methods available to all and hope that it spurs further methodological innovation within the community. 

\clearpage 
\begin{acknowledgments}
  We thank Leo Girardi for helpful discussions on theoretical stellar evolution models, Elena Sacchi for providing her resolved SFH measurement for Horologium I, and Hannah Richstein for advising on the photometry of Horologium I. Support for this work was provided by the Owens Family Foundation and by NASA through grant HST-AR-17560 from the Space Telescope Science Institute, which is operated by AURA, Inc., under NASA contract NAS 5-26555. This work is based on observations made with the NASA/ESA/CSA James Webb Space Telescope. The data were obtained from the Mikulski Archive for Space Telescopes at the Space Telescope Science Institute, which is operated by the Association of Universities for Research in Astronomy, Inc., under NASA contract NAS 5-03127 for JWST. These observations are associated with program DD-ERS-1334. This research is also based in part on observations made with the NASA/ESA Hubble Space Telescope obtained from the Space Telescope Science Institute, which is operated by the Association of Universities for Research in Astronomy, Inc., under NASA contract NAS 5–26555. These observations are associated with program HST-GO-13678. This research has made use of NASA Astrophysics Data System Bibliographic Services and the NASA/IPAC Extragalactic Database (NED), which is operated by the Jet Propulsion Laboratory, California Institute of Technology, under contract with the National Aeronautics and Space Administration.
\end{acknowledgments}

%

\facilities{JWST (NIRCAM), HST (ACS)}



\software{This research made use of routines and modules from the following software packages: 
\begin{enumerate}
    \item \href{https://github.com/cgarling/StarFormationHistories.jl}{StarFormationHistories.jl}
    \item \href{https://github.com/cgarling/InitialMassFunctions.jl}{InitialMassFunctions.jl}
    \item The Julia programming language \citep{Julia}
    \item \texttt{Matplotlib} \citep{Matplotlib}
    \item Distributions.jl \citep{Distributions.jl1,Distributions.jl2}
    \item DynamicHMC.jl \citep{DynamicHMC.jl}
    \item Optim.jl \citep{Optim.jl}
    \item \href{https://github.com/Gnimuc/LBFGSB.jl}{LBFGSB.jl} \citep{Zhu1997}.
\end{enumerate}
}

\bibliographystyle{aasjournal}
\bibliography{garling_refs}



\appendix

\section{Monte Carlo CMD Simulation} \label{appendix:montecarlo}

We implement Monte Carlo (MC) simulation of CMDs on a star-by-star basis which functions similarly to the IAC-star code \citep{Aparicio2004}. In short, stellar masses are sampled from an initial mass function (IMF) and the magnitudes of these stars are interpolated at the sampled stellar masses using user-provided isochrones. We support two different models for including unresolved, non-interacting binary systems. This sampling continues until a user-set total stellar mass or luminosity is reached. We support sampling from complex stellar populations given a separate isochrone for each SSP under consideration and a list of the fractions of initial stellar mass to allocate to each SSP (e.g., $80\%$ in a 12 Gyr population and $20\%$ in a 10 Gyr population).

We support two binary system sampling methods. In the first method, a user-defined fraction of stars are chosen to have binary companions, with their binary masses sampled randomly from the same single-star IMF independently of the first star. This method, sometimes called ``independent draws'' \citep{Gennaro2018}, is meant to be a fast, simple method and has been used in some previous studies \citep[e.g.,][]{Geha2013}, but the literature on binary systems no longer favors this treatment. In our preferred binary model, we instead sample masses from a stellar-system IMF \citep[e.g., equation 17 of][]{Chabrier2003}. Then, a percentage of these systems are randomly selected to be binary systems, and a probability distribution is used to sample a binary mass ratio that determines how the system mass is apportioned between the stars. This method, known as ``correlated draws'' \citep{Gennaro2018}, enables a large degree of flexibility, as both the binary fraction and the probability distribution of binary mass ratios can be set by the user. By default we adopt a uniform binary mass ratio distribution following \cite{Goodwin2013}.

This functionality is mainly provided as a tool for forecasting and observation planning and is not core to any of our resolved SFH measurement routines. It is worth noting that the MC technique can be used to construct partial CMDs by sampling millions of stars per SSP to achieve low Poisson errors in the model Hess diagram -- this is the approach taken by StarFISH \citep{Harris2001}, the first version of \textsc{sfera} \citep{Cignoni2015,Cignoni2019}, and the IAC-star/IAC-pop/MinnIAC software stack \citep{Aparicio2004,Aparicio2009,Hidalgo2011,Monelli2016}. This method suffers from inefficiencies due to the large dynamic range of densities in different areas of the CMDs; areas representing long phases of stellar evolution (e.g., the lower MS) will be over-sampled and areas inhabited by rapidly-evolving stars (e.g., the Hertzsprung gap) will be undersampled. The smooth template construction methods developed in this work are designed to address these issues. 

\section{SFH Recovery for Distant Galaxies} \label{appendix:sfh_examples}

\begin{figure}
  \includegraphics[width=\textwidth,page=1]{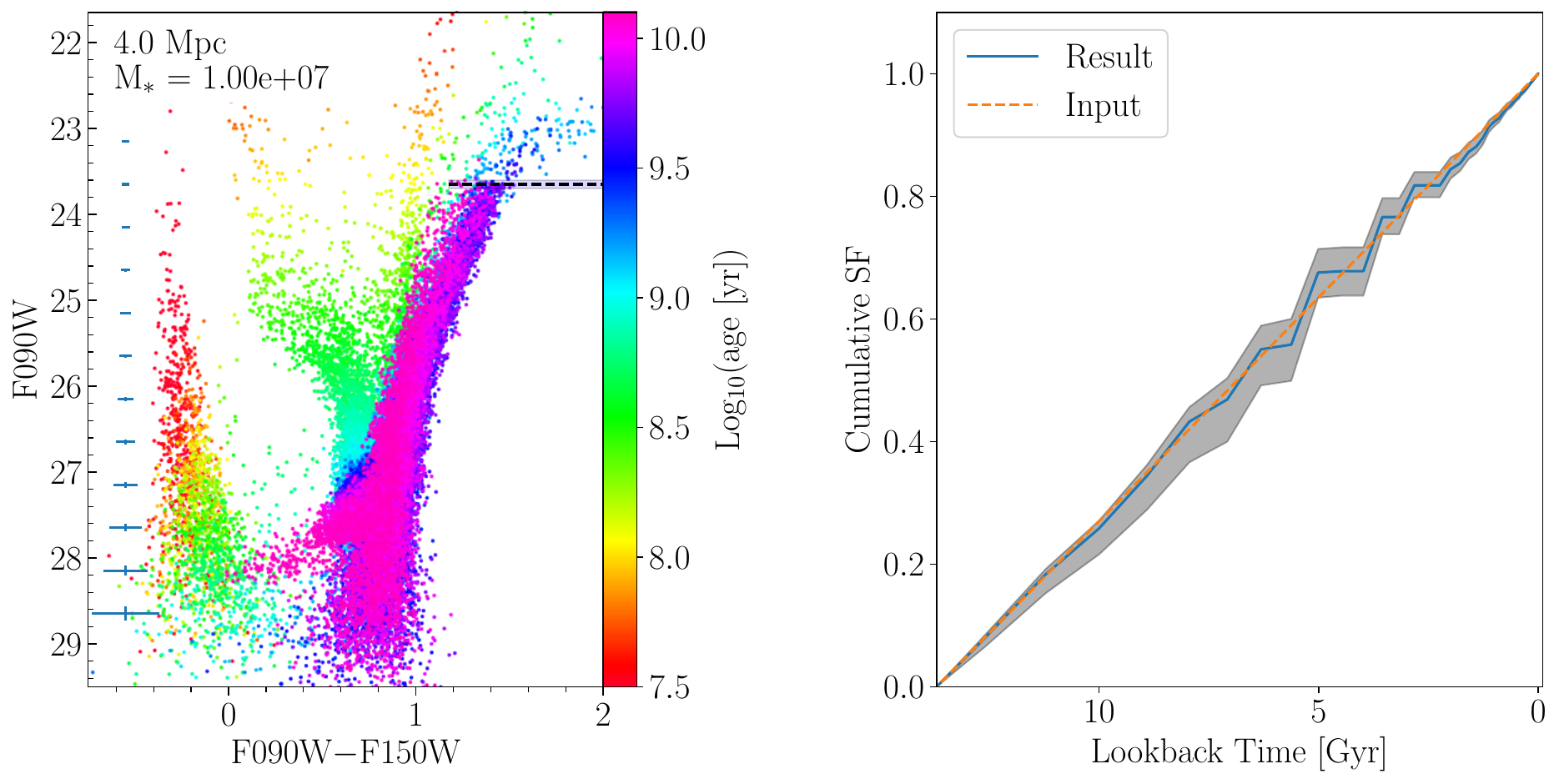}
  \includegraphics[width=\textwidth,page=1]{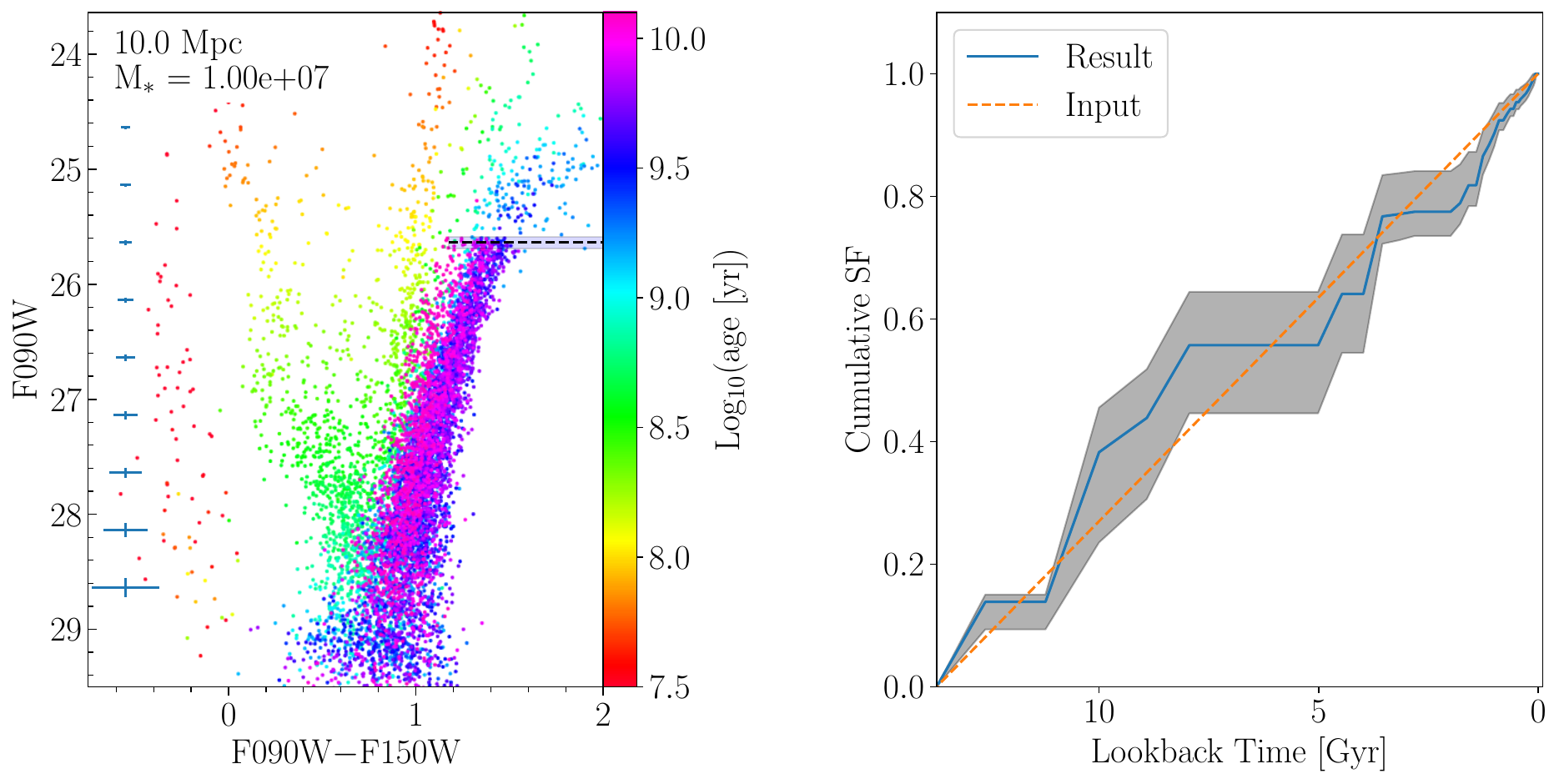}
  \caption{\emph{Left column:} Simulated CMDs of stellar populations with birth stellar masses $10^7$ M$_\odot$ with constant SFRs at distances of 4 Mpc (top) and 10 Mpc (bottom). Other modelling assumptions (e.g., photometric error and completeness functions) are the same as those used for the experiments in \S \ref{sec:synthetic_data}. Stars are colored according to their ages. The apparent TRGB magnitude in the F090W filter at the distance of the simulated populations \citep[absolute magnitude $\text{M}_\text{F090W} = -4.362$ mag;][]{Anand2024a} is shown as a dashed line. \emph{Right column:} Best-fit cumulative SFHs (solid blue line) and $68\%$ credible intervals, considering only random uncertainty, derived from the simulated CMDs in the left column. The results are consistent with the intrinsic SFH (dashed orange line). As expected, the more distant population has greater uncertainties as the shallower data resolve less of the upper main sequence, giant branch, and red clump, while the horizontal branch is altogether unresolved.}
  \label{fig:examples_appendix}
\end{figure}

In \S \ref{sec:synthetic_data} we demonstrated the performance of our SFH fitting methodology on synthetic data of a galaxy with birth stellar mass $10^7$ M$_\odot$ at a distance of 1 Mpc.
This work demonstrated our ability to robustly recover the input SFH from the simulated photometry at a distance of 1 Mpc, but there is significant interest in applying resolved SFH methods to more distant galaxies given the efficiency and sensitivity of JWST/NIRCam \citep[e.g.,][]{Bortolini2024a,Bortolini2024}. To examine the efficacy of our method at larger distances, we replicate our synthetic data experiment with synthetic photometry of galaxies at distances of 4 and 10 Mpc (distance modulii 28 and 30 mag, respectively) and show our results in Figure \ref{fig:examples_appendix}. Other than the change in the population distances, the experiment setup is otherwise the same as in \S \ref{sec:synthetic_data}.

Compared to the results for a population modelled at 1 Mpc in \S \ref{sec:synthetic_data}, the measured SFH for the population at 4 Mpc has greater random uncertainties as expected, but we are still able to recover a very accurate SFH, as the simulated photometry still reaches the old HB, the intermediate-age red clump, and the young upper MS, giving us constraining power across the full extent of cosmic time.
In comparison, the shallower CMD presented by the population modelled at 10 Mpc is lacking several of these features. The upper MS is only resolved for stars younger than 100 Myr, the red clump is highly incomplete, and the HB is entirely unresolved. The best-fit SFRs within the last $\sim1$ Gyr are quite good, as we can still leverage young blue loop and AGB stars to offset the loss of depth on the upper MS. However, the fit is noticably poorer for older ages ($>5$ Gyr) due to the unresolved HB and the shallower depth on the giant branch. This is reflected in the significantly larger random uncertainties we derive for the cumulative SFH of the population modelled at 10 Mpc. Both recovered SFHs are consistent with the input SFHs to within the derived uncertainty ranges.

This experiment illustrates that our method is appropriate for galaxies as distant as 10 Mpc, and there is no fundamental reason why it could not be applied to even greater distances. However, as the amount of information contained in the CMD is reduced for more distant galaxies, it is useful to alter some of the modelling choices to better reflect the information content of the CMD. For example, the SFHs in this experiment were fit using 71 time bins and so have fairly high time resolution -- for very distant galaxies (e.g., 20 Mpc) it is useful to reduce the number of time bins used to avoid overfitting the CMD and reduce the random uncertainties in each time bin \citep[see the discussions in][for example]{Bortolini2024a,Bortolini2024}. Modelling choices like these should be considered on a galaxy-to-galaxy basis.

\newpage

\section{Incorporating Observational Systematics} \label{appendix:obs_systematics}
Our fiducial analysis of WLM presented in \S \ref{sec:wlm} assumes fixed values for the distance modulus, MW foreground reddening, and the stellar binary fraction to mirror the analysis done with \textsc{match} by \cite{McQuinn2024} and the updated analysis with the final, public photometric catalogs \citep{Weisz2024} by \citet{Cohen2025}. However, these observational properties are never known exactly for any observed stellar population and their respective uncertainties, which fall into the category of observational systematics, should generally be propagated into the final systematic uncertainty estimate on the SFH measurement.

\begin{figure*}
  \includegraphics[width=\textwidth,page=1]{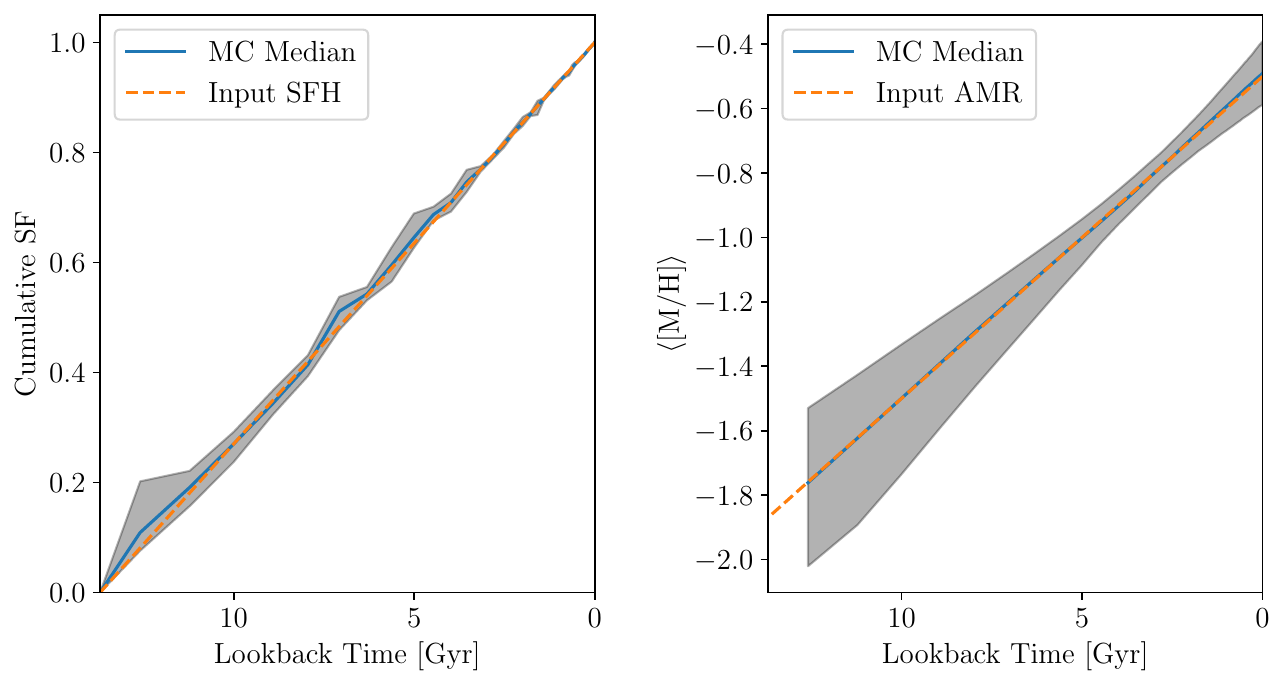}
  \caption{Effects of observational systematics (distance, foreground reddening, and stellar binary fraction) on the measurement of the cumulative SFHs (left) and AMRs (right) of synthetic stellar populations.}
  \label{fig:example_cumsfh_systematics}
\end{figure*}

To examine the effects of these observational systematics in a controlled experiment, we can simulate stellar populations with different distances, foreground reddening values, and stellar binary fractions using the methodologies developed in \S \ref{sec:synthetic_data}. We can then solve for the best-fit SFH and AMR under incorrect fiducial assumptions about these systematics in order to determine how much these observational systematics may affect the SFH measurement. We perform 1,000 such Monte Carlo experiments, sampling stellar populations with distance modulii normally distributed with $\mu=24.93$ mag, $\sigma=0.09$ mag \citep[as measured for WLM by][]{Albers2019}, stellar binary fractions uniformly sampled between 0--70$\%$, and interstellar V-band extinctions ($A_V$) uniformly distributed between 0.05--0.15 mag. We then measure the best-fit SFH for each model stellar population assuming a distance modulus of $\mu=24.93$ mag, a stellar binary fraction of $35\%$, and an interstellar V-band extinction of $A_V=0.1$ mag. These values are fixed during the fits and are, by construction, not equal to the values that were used when sampling the stellar populations. Recovering the SFH under incorrect assumptions about these observational systematics allows us to determine the extent to which these systematics affect our SFH measurements. We calculate the median and $68\%$ credible intervals for the cumulative SFHs measured from these samples and show the results in Figure \ref{fig:example_cumsfh_systematics}.

Comparing these results against Figure \ref{figure:synthetic_cumsfh}, which illustrates the cumulative SFH and AMR recovered under perfect knowledge of these observational systematics, we see that the uncertainties due to observational systematics are about twice as large as the random (statistical) uncertainties across most of the cumulative SFH. This is unsurprising as these synthetic populations consist of over 100,000 stars each and present well-sampled CMDs such that the random uncertainties are quite low. Two areas of the cumulative SFH are worth additional discussion. We see larger uncertainties due to observational systematics in the oldest two time bins (12.6 and 11.2 Gyr) which we attribute primarily to variations in the foreground extinction.
  We also see slightly larger uncertainties in the 4 -- 6 Gyr range, due primarily to variations in the distances of the populations. These synthetic populations present well-populated, tight red clump features at these ages and the systematic uncertainty in the distance mildly degrades our ability to fit this feature.

The systematic uncertainties in the AMR resulting from these observational systematics are far larger than the random uncertainties derived under perfect knowledge of these observational systematics as shown in the right panel of Figure \ref{figure:synthetic_cumsfh}. We are able to recover the AMR extremely precisely in the absence of observational systematics -- the random uncertainties on the AMR in Figure \ref{figure:synthetic_cumsfh} are so small as to be nearly imperceptible. In comparison, the uncertainties on the AMR due to observational systematics shown in Figure \ref{fig:example_cumsfh_systematics} are generally 0.1--0.2 dex. Intuitively, the AMR is more uncertain at early times -- these old stellar populations exhibit fewer metallicity-sensitive features in the CMD than younger populations, so it makes sense that the AMR should be more uncertain at these early times.

We have shown how uncertainties in observational systematics (namely the distances, foreground reddening values, and stellar binary fractions of the stellar populations) affect our ability to recover cumulative SFHs and AMRs.
The other significant systematic component in SFH measurements are theoretical systematics stemming from uncertainties in the stellar models and bolometric corrections assumed when constructing the SSP templates that are used to measure the resolved SFH (see \S \ref{subsec:templates} for a description of how stellar models are used to measure resolved SFHs). Assessing the impact of these uncertainties is non-trivial because the degree of uncertainty in the stellar models is unknown \cite[see the discussion in][]{Dolphin2012}. We leave exploration of theoretical systematics to future work.

\end{document}